\documentclass[12pt]{article}

\usepackage{authblk}
\usepackage{multibib}

\usepackage[superscript]{cite}
\usepackage[mathlines]{lineno}

\usepackage{times}

\usepackage[T1]{fontenc}
\usepackage{caption}
\usepackage{sectsty}
\sectionfont{\fontsize{13}{15}\selectfont}
\subsectionfont{\fontsize{12}{15}\selectfont}
\usepackage{nameref}
\usepackage{titlesec}

\usepackage[symbol]{footmisc}

\usepackage{amsmath}

\usepackage{amssymb}
\usepackage{lineno}
\usepackage{graphicx}
\usepackage{color}
\usepackage{bm}

\usepackage{xr}
\usepackage{abstract}

\makeatletter
\renewcommand{\maketitle}{\bgroup\setlength{\parindent}{0pt}
\begin{flushleft}
  \textbf{\@title}\\

  \@author
\end{flushleft}\egroup
}
\makeatother

\begin{document} 
\title{\noindent {\bf \large The feedback driven atomic scale Josephson microscope\\\,}}

\author[1,2]{\normalsize Samuel D. Escribano} 
\author[3]{\normalsize V\'ictor Barrena}
\author[3,4]{\normalsize David Perconte}
\author[3]{\normalsize Jose Antonio Moreno}
\author[3]{\normalsize Marta Fern\'andez Lomana}
\author[3]{\normalsize Miguel \'Agueda}
\author[3]{\normalsize Edwin Herrera}
\author[3]{\normalsize Beilun Wu}
\author[3]{\normalsize Jose Gabriel Rodrigo}
\author[5]{\normalsize Elsa Prada}
\author[3]{\normalsize Isabel Guillam\'on}
\author[1]{\normalsize Alfredo Levy Yeyati}
\author[3]{\normalsize Hermann Suderow\footnote[2]{Correspondence to samuel.diaz-escribano@weizmann.ac.il and hermann.suderow@uam.es}}

\affil[1]{\small \it Departamento de F\'isica Te\'orica de la Materia Condensada, Instituto Nicol\'as Cabrera and Condensed Matter Physics Center (IFIMAC), Universidad Aut\'onoma de Madrid, E-28049 Madrid, Spain}
\affil[2]{\small \it Department of Condensed Matter Physics, Weizmann Institute of Science, Rehovot 7610, Israel}
\affil[3]{\small \it Laboratorio de Bajas Temperaturas, Departamento de F\'isica de la Materia Condensada, Instituto Nicol\'as Cabrera and Condensed Matter Physics Center (IFIMAC), Unidad Asociada UAM-CSIC, Universidad Aut\'onoma de Madrid, E-28049 Madrid, Spain}
\affil[4]{\small \it Institut N\'eel, Univ. Grenoble Alpes, CNRS, Grenoble INP, 25 avenue des martyrs, Grenoble, 38000 France}
\affil[5]{\small \it Instituto de Ciencia de Materiales de Madrid (ICMM), Consejo Superior de Investigaciones Cient\'ificas (CSIC), E-28049 Madrid, Spain}
\renewcommand*{\Authands}{, }

\date{}

\baselineskip24pt


\maketitle 


\noindent{\bf The ultimate spatial limit to establish a Josephson coupling between two superconducting electrodes is an atomic-scale junction. The Josephson effect in such ultrasmall junctions has been used to unveil new switching dynamics, study coupling close to superconducting bound states or reveal non-reciprocal effects. However, coupling is weak and sensitivity to temperature reduces the Cooper pair current magnitude. Here we show that a feedback element induces a time-dependent bistable regime which consists of spontaneous periodic oscillations between two different Cooper pair tunneling states (corresponding to the DC and AC Josephson regimes respectively). The amplitude of the time-averaged current within the bistable regime is almost independent of temperature. By tracing the periodic oscillations in the new bistable regime as a function of the position in a Scanning Tunneling Microscope, we obtain atomic scale map of the critical current in 2H-NbSe$_2$ and find spatial modulations due to a pair density wave. Our results fundamentally improve our understanding of atomic size Josephson junctions including a feedback element in the circuit and provide a promising new route to study superconducting materials through atomic scale maps of the Josephson coupling.
}

\section*{Introduction}

Atomic-scale superconducting Josephson junctions are highly susceptible to phase fluctuations due to their small size. For small typical junctions in Al or Pb (with critical temperature T$_{\rm c}$ of 1.2 K and 7.2 K respectively), the critical current $I_{\rm c}$ is often small, in the nA range, and the Josephson coupling energy $E_{\rm J}=\frac{\Phi_0}{2\pi}I_{\rm c}$ ($\Phi_0$ being the flux quantum) can fall down to the mK range. Thus, even at temperatures well below T$_{\rm c}$, there are significant thermal fluctuations of the superconducting phase, which suppress the Josephson current.

To address the behavior of small sized Josephson junctions we start by discussing the standard resistively and capacitively shunted junction (RCSJ) model, shown in Fig.~\ref{Schematics}a (black lines)~\cite{GrossMarxDeppe}. The bias current $I_{\rm b}$ can be separated through Kirchoff's law into the sum of the supercurrent through the junction $I_{\rm c} \sin(\varphi)$ (from the first Josephson relation, $\varphi$ is the phase difference across the junction), the current through a resistance $\frac{\Phi_0}{2\pi R_S}\frac{d\varphi}{dt}$ ($R_{\rm S}$ is the resistance and the second Josephson relation is $V=\frac{\Phi_0}{2\pi}\frac{d\varphi}{dt}$), and the current through a capacitance $C_{\rm J}\frac{dV}{dt}$ (given by $\frac{\Phi_0}{2\pi} C_{\rm J}\frac{d^2\varphi}{dt^2}$). Thermal phase fluctuations induce phase difference runaway $\varphi(t)$, whose influence in the circuit can be reduced by designing the electromagnetic environment to obtain large damping, for instance through a large $RC$ impedance engineered at very small distances to the junction~\cite{Tinkham,PhysRevLett.85.170,PhysRevLett.77.3435, Joyez1999}. However, it is not always feasible to modify the local environment of a junction, for instance in Josephson junctions made with a Scanning Tunneling Microscope (STM) using a superconducting tip probing a superconducting sample.

Scanning Josephson spectroscopy (SJS) consists in obtaining maps of the near-zero voltage current with atomic precision on the surface of a superconductor. This requires a vacuum atomic size junction, which essentially fixes $R_{\rm S}$ to the vacuum impedance shunting the circuit at high frequency, and $C_{\rm J}$ to the capacitance between tip and sample. SJS has contributed to the understanding of photon assisted Cooper pair tunneling~\cite{PhysRevLett.119.147702}, tunneling at the quantum limit~\cite{Ast2016}, thermal properties of small junctions~\cite{Esat2023} or a diode effect~\cite{Trahms2023}. It has also been of help to map pair density waves (PDW) in cuprates and chalcogenides~\cite{Hamidian2016, doi:10.1126/science.abd4607}, or unveil inhomogeneous superconductivity in pnictides~\cite{Cho2019}.

A route often taken to operate systems with large thermal fluctuations is to use a delayed and time dependent feedback which couples close to the characteristic time scale of the system~\cite{DeGregorio2009,PhysRevLett.99.137205,PhysRevLett.83.3174}. However, for ultrasmall Josephson junctions, it has been recently shown that coupling to a system much slower than the Josephson time scale leads to conversion of a high frequency signal into a measurable near DC signal, and to novel dynamical behavior in between the zero voltage Josephson and the high voltage (quasiparticle) resistive regime~\cite{PhysRevLett.117.030802,PhysRevLett.128.126801,arxiv2402.09314}. Here, we show that a delayed feedback element, given by the current-voltage converter typically used in STM experiments, leads to a regime with a spontaneous oscillatory behavior, i.e. an oscillation without an external AC drive, providing a new method for SJS with minimal adjustments in the STM set-up.

\section*{Results and Discussion}
\subsection*{Feedback driven Josephson junction}

\begin{figure}
\begin{center}
	\includegraphics[width=0.85\columnwidth]{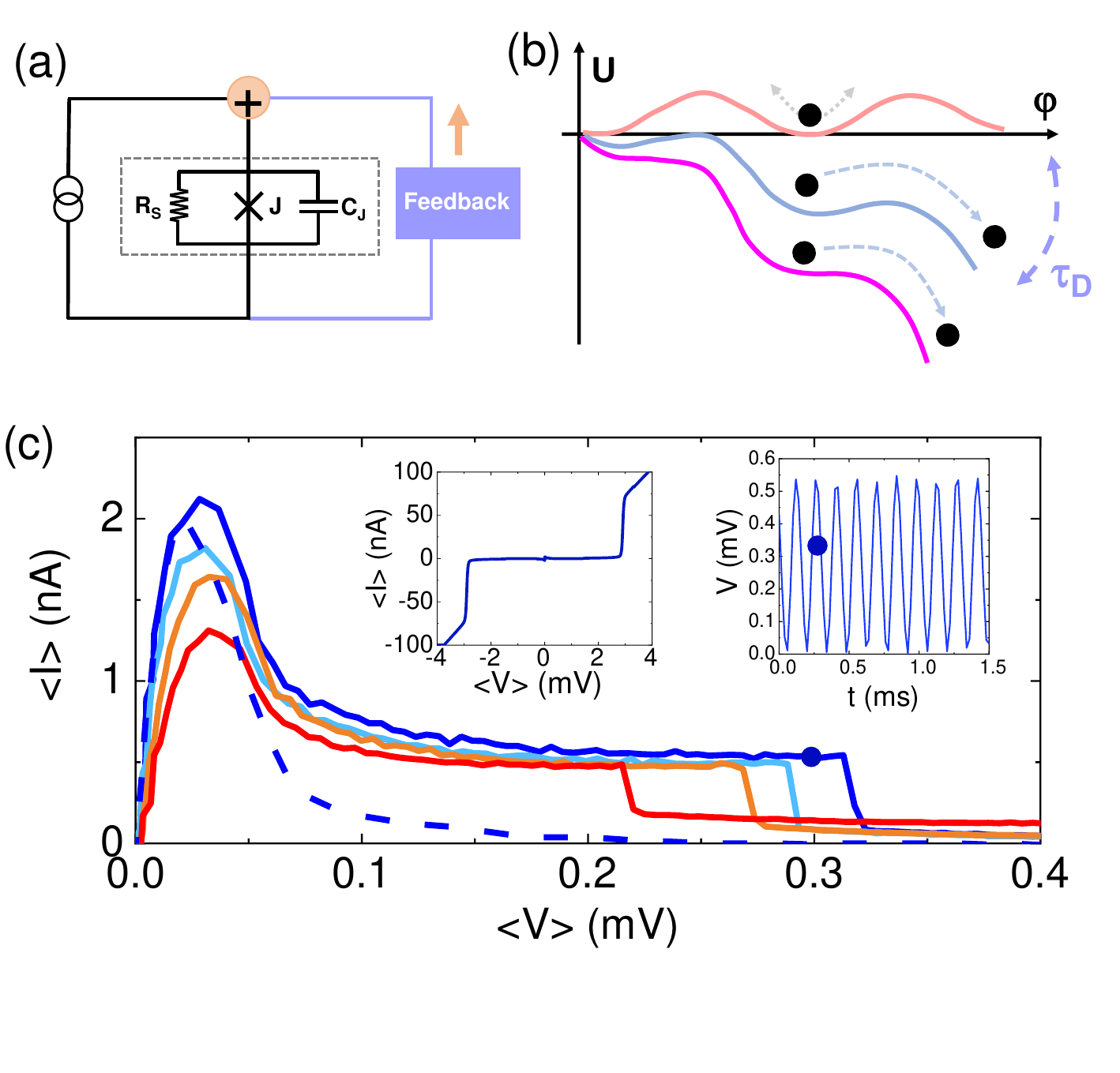}
\end{center}
\vskip -2.2cm
	\caption{\noindent {\bf \,| Feedback-driven resistively and capacitively shunted junction (RCSJ) model.} {\bf a} Schematic representation of the RCSJ model with feedback. The resistance $R_{\rm S}$, capacitance $C_{\rm J}$ and the Josephson junction J are enclosed within the grey dashed rectangle. The feedback subtracts the voltage after a delay from the actual voltage and the result is added to the circuit. {\bf b} The RCSJ model is often compared to the motion of a particle (black dots) in a periodic potential $U$ (lines) along a coordinate which represents the phase difference across the Josephson junction $\varphi$. At low bias (salmon), thermal excitation induces phase fluctuations (grey dashed arrow, the black dots schematically represents the value of $\varphi$ at a certain time). By applying a bias to the junction, the periodic potential $U$ is tilted downwards, as illustrated by the blue and magenta lines, causing the particle to slip along the potential. Incorporating a feedback mechanism with a time constant $\tau_{\rm D}$ periodically modifies the tilt in $U$ (violet dashed line). {\bf c} Time-averaged current $\langle I\rangle$ vs time-averaged voltage $\langle V\rangle$ measured in a STM junction between a Pb tip and a Pb sample at different temperatures: $T=0.15$~K (solid blue), $T=1$~K (cyan), $T=2$~K (orange) and $T=3$~K (red). Tunneling conductance is $G\approx 0.4 G_0$, being $G_0$ the conductance quantum. Further details of the experiment are given in Methods Section 1. These curves are a zoom around zero voltage of the $\langle I\rangle -\langle V\rangle $ curve shown in the upper left inset. Dashed blue line are data taken without a feedback (see also Methods Section 1), at $T\approx 0.15$~K and $G\approx 0.4 G_0$. The upper right inset shows the time-dependent trace of the voltage at the point in the current-voltage curve marked by a blue point in the main panel. We show, for clarity, only curves obtained by ramping up the bias.}
	\label{Schematics}
	\end{figure}

To understand the effect of a feedback coupling on a Josephson junction, we recall that the phase $\varphi$ behaves as the position of a particle in a one dimensional potential $U$. We can write $U(\varphi)=\int d\varphi \left(I_{\rm b}-I_{\rm c} \sin(\varphi)\right)$. $U$ has the form of a tilted washboard potential as shown in Fig.~\ref{Schematics}b~\cite{doi:10.1002/3527603646.ch2}. With a small $I_{\rm b}$ (orange curve in Fig.~\ref{Schematics}b), the phase difference $\varphi$ remains around a potential minimum, resulting in zero voltage across the junction. As the bias is increased (blue and magenta curves), the phase rolls down the potential, leading to a non-zero voltage (often at this point Cooper pair tunneling ceases and the junction jumps into quasiparticle tunneling). A feedback adds a time delayed current to the actual Josephson current and periodically inhibits the evolution of the phase with time by reducing the tilt, as schematically shown in Fig.\,\ref{Schematics}a,b. As we show below, this periodic behavior induced by the feedback leads to a regime which presents spontaneous oscillations at the time constant $\tau_D$ of the circuit and a time averaged current which is almost independent of temperature.

In Fig.~\ref{Schematics}c, we compare the time-averaged current vs the time-averaged voltage, $\langle I\rangle-\langle V\rangle$, for a Josephson junction formed between a Pb STM tip and a Pb sample (blue line) when a feedback element is integrated into the circuit, to the result obtained without a feedback element (dashed blue line, the experiment is described below and in Methods Section 1). We observe a regime with a sustained current in the circuit with a feedback over a bias range that is considerably larger than in the circuit without a feedback. Whereas the behavior close to zero bias is highly temperature dependent (following approximately the same behavior as in the circuit without a feedback), the current in the regime induced by the feedback is almost independent of temperature (see Fig.\,\ref{Schematics}c). Importantly (upper right inset of Fig.\,\ref{Schematics}c), in this regime there are temporal oscillations which can be used as a new probe for microscopy, as we show below.

\begin{figure}
\begin{center}
	\includegraphics[width=\textwidth]{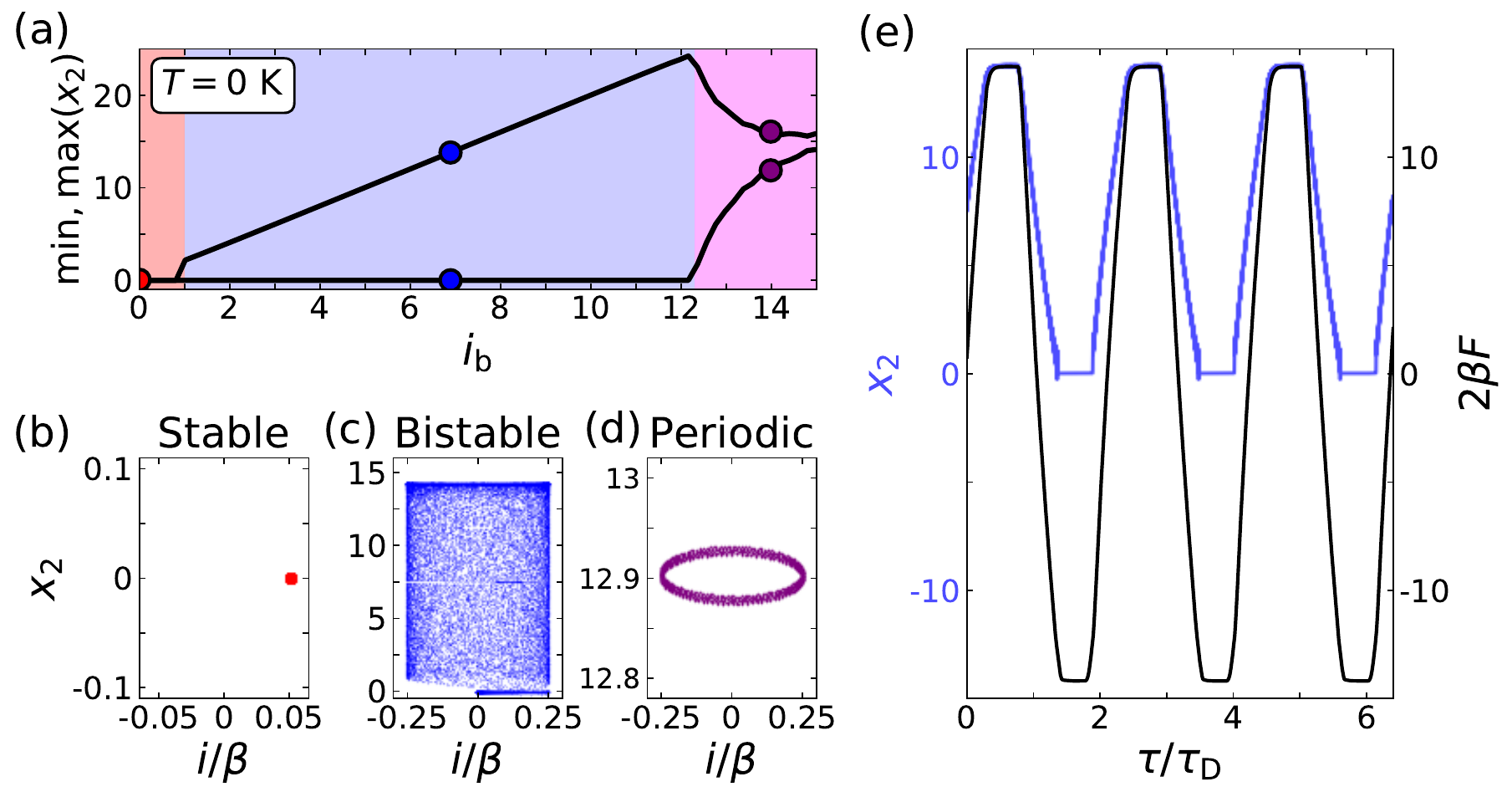}
\end{center}
	\caption{\noindent {\bf  \,| Feedback driven RCSJ regimes.} {\bf a} The minimum and maximum values of $x_2$ (proportional to the junction voltage) as a function of the reduced bias current $i_{\rm b}=I_{\rm b}/I_{\rm c}$. We identify three different regimes, shadowed with orange, blue, and purple backgrounds. {\bf b-d} Evolution of the current [$i=\sin(x_1)$] and the voltage ($x_2$) through the junction at different times. The color in each panel represents schematically the regime, following the color of the dots in {\bf a}. In {\bf b} the junction quickly reaches a stable state with a time-independent finite current and zero voltage (stable regime). In {\bf c} the junction leaves the zero voltage state and oscillates between a zero and a finite voltage state (bistable regime). Oscillations are spontaneous, i.e. without the action of an external AC drive. In {\bf d} both the current and the voltage oscillate at frequency $\frac{2\pi}{\tau_{\rm J}}$ (periodic regime), but the current averages to zero while the voltage averages to a finite value. {\bf e} $x_2$ vs time (blue line) together with the feedback term multiplied by $2\beta$ (black line), see Eq.~\eqref{RCSJFeed}. Parameters in these simulations are $T=0$ K, $\beta=2$, $\tau_{\rm D}=10^5$ and $\Delta\tau=6\cdot10^3$ (both $\tau_{\rm D}$ and $\Delta\tau$ in units of $\tau_{\rm J}$).}
	\label{FigOscillations}
	\end{figure}

To gain a deeper understanding, let us start by establishing the feedback RCSJ model at zero temperature. We can write
\begin{equation}
    \frac{dx_2}{d\tau}= \frac{i_{\rm b}-\sin(x_1)-x_2}{\beta} + F(\tau,\tau_{\rm D})
\label{RCSJFeed},\\
\end{equation}
where the feedback $F(\tau,\tau_{\rm D})$ is given by
\begin{equation}
    F(\tau,\tau_{\rm D})=-\frac{1}{2\beta}\int_{\tau}^{\tau+\Delta\tau}\frac{1}{\Delta\tau}\left[x_2(\tau'-\tau_{\rm D})-x_2(\tau')\right]d\tau'.\\
\label{Feedback}
\end{equation}
Here, $x_1\equiv \varphi$ and $x_2\equiv \frac{d\varphi}{d\tau}$ are the dimensionless variables, the reduced time is $\tau=\frac{t}{\tau_{\rm J}}$, with $\tau_{\rm J}=\frac{\Phi_0}{2\pi I_{\rm c}R_{\rm S}}$ the characteristic Josephson time (and $\omega_{\rm J}=\frac{2\pi}{\tau_{\rm J}}$ the Josephson frequency), the reduced bias current is $i_{\rm b}\equiv I_{\rm b}/I_{\rm c}$, and $\beta=\frac{2\pi I_{\rm c}R_{\rm S}^2 C_{\rm J}}{\Phi_0}$ is the McCumber parameter. Eq.\,\ref{RCSJFeed} corresponds to the conventional RCSJ model plus the feedback term $F(\tau,\tau_{\rm D})$. $\Delta\tau$ is the bandwidth of the measurement circuit. The behavior of the solutions to the new coupled equations $\frac{dx_1}{d\tau}=x_2$ and Eq.~\eqref{RCSJFeed} is shown in Fig.~\ref{FigOscillations}.

In Fig.~\ref{FigOscillations}a-d we show the evolution of the voltage $x_2$ as function of $i_{\rm b}$. We identify three distinct regimes.  For $i_{\rm b}\le 1$ (orange in Fig.~\ref{FigOscillations}a and Fig.~\ref{FigOscillations}b) we find zero voltage, $x_2=0$, and a finite Josephson current $i=i_{\rm b}$. In this regime, Eq.\,\ref{RCSJFeed} is the usual DC Josephson junction equation $i_{\rm b}=sin(x_1)$. The most interesting regime is the bistable regime, which sets in when $i_{\rm b}\simeq 1$. Then, $x_2(\tau)\neq0$. The current through the junction $i$ (current through the middle branch within the dashed rectangle in Fig.~\ref{Schematics}a) first increases and then oscillates at a time scale which is of order of $\tau_{\rm J}$, as shown in Fig.~\ref{FigOscillations}c, leading to a finite voltage $x_2(\tau)\neq0$. However, just when $i_{\rm b}$ is increased above one, $x_2(\tau-\tau_{\rm D})$ is still zero, which leads to a non-zero feedback term $F(\tau,\tau_{\rm D})$. As $x_2(\tau)$ increases with time (blue line in Fig.~\ref{FigOscillations}e), the feedback $F(\tau,\tau_{\rm D})$ increases (black line in Fig.~\ref{FigOscillations}e). But $x_2(\tau-\tau_{\rm D})$ increases too and thus at some point the feedback $F(\tau,\tau_{\rm D})$ decreases again. Then, $x_2(\tau)$ decreases until $x_2(\tau)=0$ again. $x_2(\tau-\tau_{\rm D})$ decreases and in this process the feedback changes sign. Having now a negative feedback leads to increasing $x_2(\tau)\neq0$, and the process starts again.

When $i_{\rm b}\simeq 1$, the feedback exactly compensates the additional current above the critical value in Eq.\,\ref{RCSJFeed} within a short amount of time and the oscillations in $x_2(\tau)$ and $F(\tau,\tau_{\rm D})$ have essentially a square shape. But when $i_{\rm b}$ increases, the compensation sets in only gradually. For example, in Fig.~\ref{FigOscillations}c we are approximately in the middle of the bistable regime. The time interval where the feedback and $x_2$ vary occupies approximately half a cycle ($0.5\tau_D$). This interval increases with $i_{\rm b}$, until the feedback no longer compensates the appearance of a finite voltage and the oscillations in $x_2(\tau)$ and $F(\tau,\tau_{\rm D})$ are triangular like. Then, we enter another regime. The third regime is similar to the conventional AC periodic Josephson regime, characterized by a finite time-averaged voltage $x_2=i_{\rm b}$ and a zero time-averaged current (purple in Fig.~\ref{FigOscillations}a and Fig.~\ref{FigOscillations}d).

The bistable regime is characterized by an oscillating behavior which appears without a drive, at a the feedback time constant $\tau_D$, and is independent of the Josephson frequency $\tau_J$. Furthermore, the current and voltage are determined by the time dependent behavior, rather than by their sensitivity to temperature.

\begin{figure}
\begin{center}
	\includegraphics[width=0.9\columnwidth]{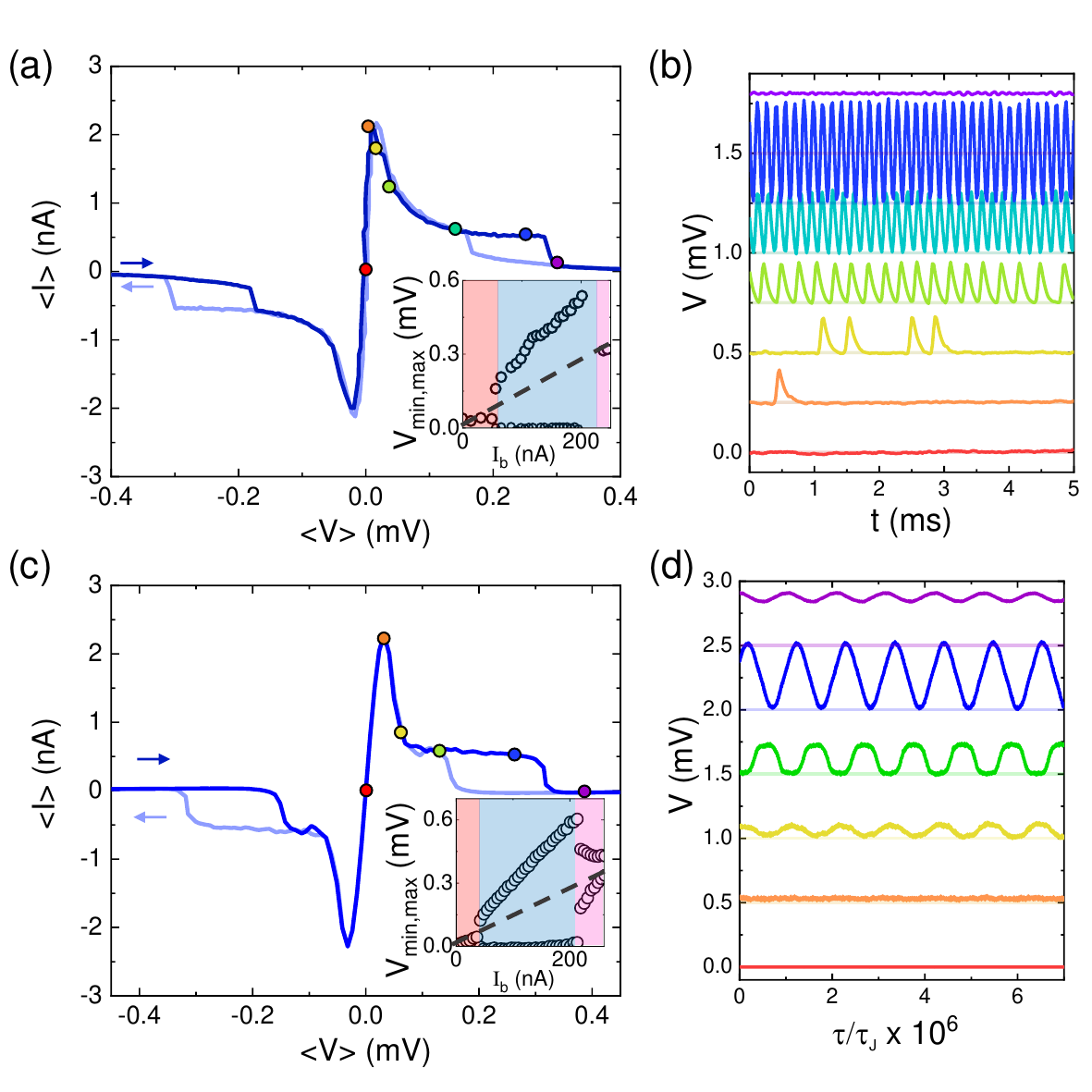}
\end{center}
\vskip -1cm
	\caption{\noindent {\bf  \,| Feedback driven Josephson effect.} {\bf a} Time-averaged tunneling current vs time-averaged voltage, $\langle I\rangle -\langle V\rangle $, near zero bias (solid blue lines) for a STM atomic size Josephson junction made between two Pb electrodes. The blue arrows indicate the direction of the ramp. Colored dots represent the locations at which we stop the ramp and make time-dependent measurements, shown in {\bf b}. The junction conductance $G$ is $\sim$0.4$G_0$, being $G_0=\frac{2e^2}{h}$ the conductance quantum and $T=$150~mK. The bottom right inset shows the maximum and minimum of the voltage vs $I_{\rm b}$. {\bf b} Voltage vs time at the colored dots marked in the main panel of {\bf a}. The curves are vertically shifted by 0.25~mV for clarity. {\bf c} Time-averaged current vs time-averaged voltage, $\langle I\rangle -\langle V\rangle $, obtained from the model described in the text. The arrows provide the direction of the ramp. The bottom right inset shows the maximum and minimum of the voltage vs $I_{\rm b}$. {\bf d} Voltage as a function of time for the colored dots marked in {\bf c}. Curves are shifted for clarity by 0.5~mV. Note that $\tau$ is in units of $\tau_{\rm J}\approx 10^{-9}$~s, see text and Methods (Sections 2,3) for further details. Parameters in the simulations are $T=1.2$~K, $\beta=2$, $\tau_{\rm D}=10^5$ and $\Delta\tau=6\cdot10^3$ (both $\tau_{\rm D}$ and $\Delta\tau$ in units of $\tau_{\rm J}$).}
	\label{Josephson}
	\end{figure}

We now compare theory and experiment at a finite temperature. In Fig.~\ref{Josephson}a, we show a time-averaged current-voltage characteristic taken on an atomic size tunnel junction between a tip and a sample of Pb ($T_{\rm c}^{\rm (Pb)}=7.2$~K, and $\Delta_{\rm Pb}=1.37$~meV, $\tau_{\rm J}\approx 10^{-9}$ s). Starting from zero bias, we observe that the average current increases, reaches a peak, and then decreases and saturates at a plateau, eventually dropping to zero at higher bias. The time dependence of the voltage for different bias (colored points in Fig.~\ref{Josephson}a) is shown in Fig.~\ref{Josephson}b and the minimum and maximum values in the inset of Fig.~\ref{Josephson}a (we use $\Delta\tau\sim 10^{-5}$ s for the temporal range or bandwidth of the experimental set-up, see also Methods Section 1). We observe a bistable regime (orange to blue curves) where the junction oscillates periodically, approximately every $0.15$~ms, between zero and a finite voltage. We note that the oscillation appears spontaneously, without the action of any AC drive. In Fig.~\ref{Josephson}c we show the time-averaged current vs the time-averaged voltage resulting from our simulations. The voltage vs time is shown in Fig.~\ref{Josephson}d and the minimum and maximum values in the inset of Fig.~\ref{Josephson}c. We see that experiment and calculations are mostly identical. The shape of the time-averaged current vs time-averaged voltage curve is similar, the hysteresis has a similar shape and, most importantly, the oscillatory behavior in the bistable regime occurs in experiment as well as in simulations.

The occurrence of the bistable regime depends on the direction for ramping the bias (light and dark blue lines in Fig.~\ref{Josephson}a,c show the different ramping directions). The model shows that, when decreasing $i_{\rm b}$, $x_2(\tau-\tau_{\rm D})\neq0$ and the feedback $F(\tau,\tau_{\rm D})$ (Eq.\,\ref{Feedback}) remains at a small value. Thus, a smaller $i_{\rm b}$ is needed to reach the bistable regime.

This  behavior is completely different from the known behavior of hysteretic Josephson junctions~\cite{GrossMarxDeppe}. Instead of switching out of the Josephson Cooper pair tunneling regime into a quasiparticle branch, entry in and exit out of the bistable regime is between two different regimes of Cooper pair tunneling. It occurs between states that are similar to the DC (finite current and zero voltage at the junction) and AC (finite voltage and a rapidly oscillating current at the junction) Josephson effects of an isolated Josephson junction.

Increasing the temperature reduces both the bistable regime as well as the size of the hysteretic behavior (Methods Section 2 and 3), but leads, as shown in Supplementary Fig.\,\ref{Josephson_T_R}, to a temperature independent value of the time-averaged current in the bistable regime.

A detailed analysis of the different parameters governing the feedback RCSJ model (Supplementary Fig.\,\ref{Parameters} and Methods Section 4) shows that the appearance of the bistable regime for different time constants $\tau_{\rm D}$ goes all the way down to $\tau_{\rm D}$ being one or two orders of magnitude larger than $\tau_{\rm J}$. Furthermore, there is a large range of $\beta>1$ where the bistable regime can be observed. On the other hand, $\Delta\tau$ should be smaller than $\tau_{\rm D}$ but below $\Delta\tau\approx 0.1 \tau_{\rm D}$, the integration does no longer provide a sufficiently large feedback term. In the experiment, we have observed the bistable regime in Pb-NbSe$_2$ and Al-Al atomic size junctions (Supplementary Figs.\,\ref{PbNbSe2},\ref{Al} and Supplementary Information Section 1). We note that $\tau_D$ is controlled by an $RC$ filter in the amplifying circuit and discuss the detailed dependence on the circuit parameters in Methods (Section 1 and Supplementary Fig.\,\ref{Scheme}).

Note that there is a discrepancy between the symmetric peaks obtained in the voltage as a function of time in the bistable regime of our simulations, and the asymmetric peaks observed in the experiment (yellow and green curves in Fig.~\ref{Josephson}b and Fig.~\ref{Josephson}d). We also note that the oscillations in the model are always periodic, whereas we observe a stochastic instability to enter the bistable regime in the experiment (see Supplementary Information Section 2). We do not have a clear explanation for these discrepancies. The asymmetry might be related to asymmetries in the circuit. For instance, different resistance and capacitance values in the two branches connecting the junction (Fig.\,\ref{Schematics}a), which are not included in our feedback RCSJ model. The stochastic peaks could be related to excitations, as photons, driving the junction into the bistable regime. The other essential features, such as the existence of a bistable regime and its accompanying hysteretic behavior, are accurately captured by the feedback RCSJ model.

\subsection*{Critical current from the bistable regime in feedback driven ultra small Josephson junctions}

\begin{figure}
	\includegraphics[width=\textwidth]{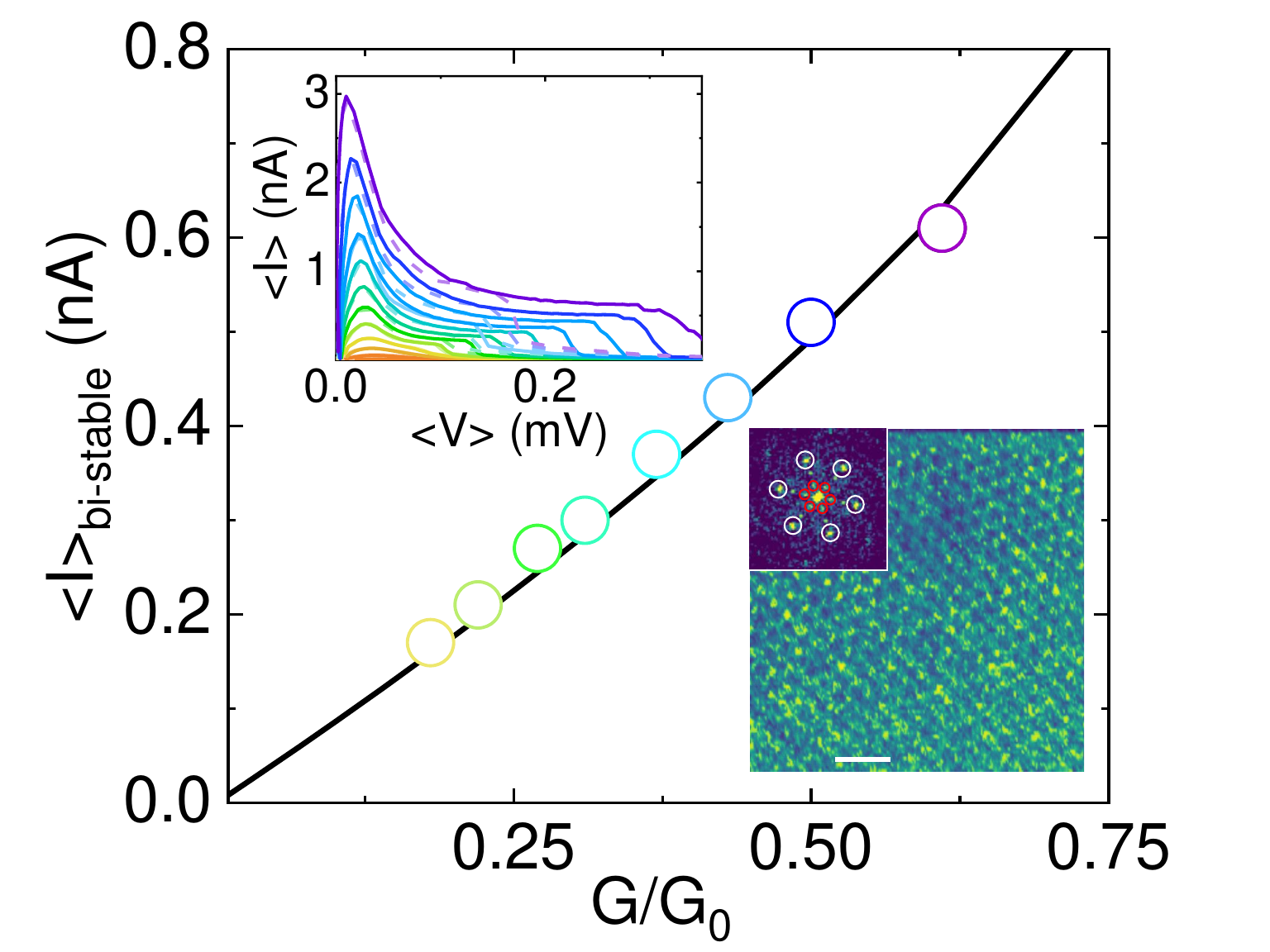}
	\caption{\noindent {\bf  \,| Current in the bistable regime vs conductance.}  Time-averaged tunneling current $\langle I\rangle_{bistable}$ just before the loss of the bistable regime as a function of the conductance $G$ normalized to the conductance quantum $G_0$, for STM Josephson junctions made with tip and sample of Pb. The colored circles correspond to the experimental results whereas the line corresponds to the expression $I_{\rm c}\frac{\Delta\tau}{4\tau_{\rm D}}$. 
    $I_{\rm c}$ is determined as a function of $G$ following Ref.\,\cite{PhysRevB.51.3743}. In the upper left inset we show the time-averaged tunneling current $\left<I\right>$ vs the time-averaged voltage $\left<V\right>$. Each line corresponds to a different tunneling conductance, following the color code of the circles in the main panel ($G=0.61G_0$ (violet), 0.50$G_0$ (blue), 0.43$G_0$ (light blue), 0.37$G_0$ (cyan), 0.31$G_0$ (light cyan), 0.27$G_0$ (green), 0.22$G_0$ (light green), and 0.18$G_0$ (yellow)). Dashed lines in the inset are for decreasing bias. In the lower right inset we show a map of the amplitude of the oscillatory signal in the bistable regime showing atomic resolution in 2H-NbSe$_2$. The amplitude of the oscillatory signal improves significantly the signal to noise ratio and increases with $I_c$. White scale bar is 1 nm long. In the upper left corner of the inset we show the Fourier transform, where we can identify the Bragg peaks of the hexagonal Se lattice (white circles) and the Bragg peaks due to the PDW (red circles, more details in Supplementary Figure 6 and in the Supplementary Information Section 4).} 
	\label{Fig_OscDepen} 
	\end{figure}

From the feedback RCSJ model at a finite temperature, we find that $\left<I\right>=I_{\rm c}\frac{\Delta\tau}{4\tau_{\rm D}}$ in the bistable regime, just before the transition to the periodic regime. This is in sharp contrast to the value of the current close to zero voltage, for $i_{\rm b}\approx 1$, which inversely proportional to temperature, $\sim T^{-1}$, or with the derivative of the current near zero voltage, which decreases with temperature as $\sim T^{-2}$ (see Supplementary Information Section 3). For the experiment shown in Fig.~\ref{Josephson}a, we find $I_{\rm c} \sim50$~nA, which is very similar to the estimation provided by the Ambegaokar-Baratoff equation $I_{\rm c}=\frac{G}{G_0}\frac{e}{2\hbar}\Delta$ for symmetric junctions at zero temperature~\cite{PhysRevLett.10.486} ($G$ being the conductance and $G_0$ the quantum of conductance, and $\Delta$ the superconducting gap, we use $G\approx 0.4G_0$ and $\Delta_{\rm Pb}=1.37$~meV). By taking $\Delta\tau=0.06 \tau_{\rm D}$ and $I_{\rm c} \sim50$~nA we can  follow closely the dependence of $I_{\rm c}$ on $G$, as shown in Fig.~\ref{Fig_OscDepen}, finding that the experiment (dots) follows the detailed calculations of $I_{\rm c}$ vs $G$ of Ref.\,\cite{PhysRevB.51.3743} (line). Thus, the bistable regime provides a quantitative measurement of the critical current $I_{\rm c}$ of the junction.

\subsection*{Pair density wave in 2H-NbSe$_2$ observed on atomic resolution maps of the oscillatory component}

We have observed the bistable regime in a Josephson junction with a Pb tip and a superconducting 2H-NbSe$_2$ sample, and used the bistable regime to obtain a considerably improved measurement of the atomic scale variation of the Josephson critical current in 2H-NbSe$_2$, as compared to the measurement of the current or the tunneling conductance close to zero bias used in previous SJS experiments (details in Methods Section 5 and in the Supplementary Information Sections 4 and 5). We have followed the amplitude of the spontaneous oscillatory signal as a function of the position. The amplitude of the oscillation increases with bias, as shown in the insets of Fig.\,3a,c, until the system leaves the bistable regime. It is proportional to the bias range in which the bistable regime is found and to the critical current $I_c$ , as shown in the inset of Fig.\,4. Measuring the amplitude of the oscillation in the bistable regime, we benefit from the improved detection of an oscillatory signal at a well-defined frequency and achieve a considerable improvement of the signal to noise ratio. 2H-NbSe$_2$ has a charge density wave (CDW) below T$_{CDW}=$33 K which coexists with superconductivity below T$_c=$7.2 K. 2H-NbSe$_2$ presents atomic size modulations of the superconducting order parameter\cite{Guillamon2008}. These modulations are linked to the charge modulations of the CDW through a pair density wave (PDW)\,\cite{doi:10.1126/science.abd4607}. The PDW consists of spatial modulations of a coherent pair density\,\cite{doi:10.1146/annurev-conmatphys-031119-050711}. The PDW has been observed in cuprates, pnictides and heavy fermions, and is believed to cover the link between charge order and Cooper pair formation that prevails on the phase diagram of high T$_c$ superconductors\,\cite{Hamidian2016,Edkins976,doi:10.1073/pnas.2206481119,Zhao2023,Liu2023,Aishwarya2023,Gu2023}. As shown in the lower right inset of Fig.~\ref{Fig_OscDepen}, we observe the PDW by mapping the oscillatory signal in the bistable regime.

The newly uncovered bistable regime is a considerable addition on Josephson probes, of particular interest in Josephson junctions including a topological superconductor~\cite{PhysRevLett.103.197002, PhysRevLett.105.077001, PhysRevB.67.220504, PhysRevLett.105.177002, PhysRevLett.105.097001, PhysRevB.96.184501, PhysRevLett.119.197002, PhysRevB.100.045301, Zazunov2018, AGP-2018:123,Aishwarya2023,Gu2023} or a chain of magnetic adatoms \cite{Nadj-Perge2014, Schneider2023}. In those cases, Cooper pair tunneling between the s-wave state of the tip and p-wave state of the sample is expected to occur through higher-order processes with a significant reduction of the usual $I=I_c sin(\varphi)$ current phase relation~\cite{Zazunov2018}. Mapping the bistable regime with atomic registry in these systems should provide radically new information from the measurement of an oscillating signal that arises without an AC drive, and the new phenomena arising from the Josephson relation modified by the feedback. The bistable regime opens thus new avenues to study unconventional superconductivity.

In conclusion, we have unveiled a bistable regime of Josephson junctions governed by a feedback action and providing Josephson characteristics in a frequency range six orders of magnitude below the Josephson frequency. The oscillating behavior is a defining feature of ultra small size feedback driven Josephson junctions and can be used to map the Josephson effect down to atomic scale through a Josephson dynamic behavior which is very different from dynamics of usual tunnel junctions.

\clearpage

\section*{Methods}

\section{Measurement of the Josephson current with STM}

We use a STM set-up in a dilution refrigerator with superconducting tip and sample. The tips are always prepared and cleaned through repeated indentation inside a pad of the same material as the tip. We indent the tip in a controlled way and follow the conductance steps characteristic of atomic-size junctions~\cite{Rodrigo2004, Suderow2011}. The same preparation method is used for tip and sample when measuring Pb-Pb, Pb-2H-NbSe$_2$, and Al-Al junctions. The STM set-up, the software used and aspects of the cryogenics are described in Refs.~\cite{Suderow2011, doi:10.1063/5.0059394, Galvis2015, MONTOYA2019e00058, doi:10.1063/1.2432410, doi:10.1063/5.0064511}. The 2H-NbSe$_2$ sample has been cleaved in-situ at low temperatures.

In our setup (Supplementary Fig.\,\ref{Scheme}), we employ a circuit featuring an operational amplifier for the feedback element, which is commonly utilized for SJS measurements~\cite{Hamidian2016, doi:10.1126/science.abd4607, Cho2019, PhysRevLett.119.147702, Huang2020, Trahms2023, PhysRevB.64.212506, PhysRevLett.87.097004,PhysRevB.80.144506,PhysRevB.93.161115,Rodrigo04,10.1063/1.122654,Rodrigo2006PhysC,GUILLAMON2008537,PhysRevB.93.020504,KOHEN200518,PhysRevB.78.140507,Proslier_2006,Edkins976,Ast2016,Joo2019,doi:10.1073/pnas.2206481119,Liu2021,PhysRevLett.130.177002}. There is nevertheless an important difference between our circuit and similar ones. Often, the resistor $R$ is chosen within the range of $1$ to $10$~M$\Omega$  (amplification between $10^6$ to $10^7$) and an additional filtered amplification stage is employed to achieve current-to-voltage conversions in the $10^8$, $10^9$ or higher ranges. We do not use the latter filter in the amplification stage, because it can eliminate or reduce any oscillatory behavior present in the junction like the bistable regime discussed in this work.

In absence of such an additional filtering stage, the current flowing through the atomic-size tunnel junction is directly measured through the operational amplifier in a current-to-voltage configuration. In principle, the junction is voltage biased. But, in our setup we use a voltage source with a 10~k$\Omega$ resistor connected in series. The voltage drop on this resistor is subtracted from the applied voltage to obtain the time-averaged $\left<I\right>-\left<V\right>$ curves. We determine the voltage as a function of time by multiplying the output of the operational amplifier by the resistor in parallel to the amplifier $R$ and subtracting an offset.

The role of the feedback of the operational amplifier is to make sure that the current flows through the resistor $R$ and reaches ground, thereby simultaneously nullifying the voltage difference at the amplifier's two inputs. However, the feedback's action is not immediate but rather delayed by a certain time constant $\tau_{\rm D}$, limited by the junction's resistance and the capacitance $C$. This delay leads to the spontaneous oscillatory behavior discussed here.

The capacitance $C$ in our circuit is mainly due to the capacitance in parallel to the resistor in Supplementary Fig.\,\ref{Scheme}, which mostly consists of the wiring to ground. This can be significant due to the typical long wiring in a dilution refrigerator setup~\cite{doi:10.1063/5.0064511, doi:10.1063/5.0059394}. We observe the oscillatory behavior when $C_{\rm wire}\simeq 0.3$~nF. In the Al-Al experiment (discussed in the Supplementary Information Section 1), we use a capacitor of $C\approx 2$~nF and get an approximately ten times larger $\tau_{\rm D}$. When the setup has a capacitance of $\sim 10$~nF, $\tau_{\rm D}$ further increases and the bistable regime is lost, resulting in the Josephson behavior reported in previous works, discussed in more detail in the Supplementary Information Sections 1,6 and shown in Fig.\ref{Schematics}b of the main text (blue dashed line).

In our experiment, taking $\beta=\frac{2\pi}{\Phi_0}I_{\rm c}R_{\rm S}^2 C_{\rm J}=4$, we estimate that $C_{\rm J}$ is on the order of, or below, a few fF. The capacitance of tunnel junctions made with STM have been measured and calculated for typical tip shapes and sample-tip distances, yielding values in the range of tens of fF~\cite{DEVOOGD201761,Esat2023}. These capacitance values indicate that the tip can be modeled by a sphere with an effective radius close to $1$~$\mu$m. This suggests that the junction capacitance $C_{\rm J}$ is primarily determined by the geometry of the tip at a distance slightly below that value, typically in the range of a few tens to hundreds of nm.

We use the OPA 111 amplifier in our setup, featuring a low bias current~\cite{OPA111}. In the operational amplifier, two field effect transistors (FET) are connected to the input and to the power supply. The inputs ($+$ and $-$ in Supplementary Fig.\,\ref{Scheme}) are connected to the gate of each FET while the power supply is connected to the source and the output to the drain of each FET. Before reaching the output, the signal is amplified through a low-noise cascode. As the gain is very large in the used circuit, we expect the settling time to be of order of the RC time of the circuit.

\section{Temperature dependence in the feedback-driven RCSJ model}

The entry and exit behavior between different regimes is thermally driven, as shown in the Supplementary Fig.\,\ref{Histo} and discussed in the Supplementary Information Section 3. SJS in the bistable regime provides an exit behavior that can be traced as a function of the position. The full loss of Cooper pair transport between tip and sample occurs when leaving the bistable regime by increasing the bias. Entering and leaving the bistable regime occurs at a bias which is far above the usual bias at which a thermally smeared Josephson current is observed (Fig.\,\ref{Josephson}). As shown in the Supplementary Information Section 3, this eventually allows identifying and studying non-reciprocal (diode) Josephson behavior. Furthermore, the value of the current in the bistable regime is given by the critical current, but the exit point depends on the shape of the current phase relation. As we show in Supplementary Fig.\,\ref{Switch} and discuss in the Supplementary Information Section 5, the exit out of and entry into the bistable regime is different when taking a current phase relation deviating from $I=I_c sin(\varphi)$. In the bistable regime we can make atomic size maps of the exit behavior by tracing directly the oscillatory signal as a function of the position. We observe (Supplementary Fig.\,\ref{SwitchMap}) spatial modulations at the PDW wavevector.

We follow the feedback-driven RCSJ model, see Refs.~\cite{LiSen2011, noise_RK}. By using the Kirchhoff equations on the circuit of Fig.~\ref{Schematics}a and introducing the feedback term and thermal noise, we obtain the set of differential equations
\begin{align}
\label{Eq1b} 
V &=\frac{\Phi_0}{2\pi}\frac{d\varphi}{dt}, \\
I_{\rm b}-k\frac{1}{\Delta t}\int_{t}^{t+\Delta t}\left[V(t'-t_{\rm D})-V(t')\right]dt' &=& \nonumber \\ 
I_{\rm c}\sin{\varphi} +\frac{V}{R_{\rm S}}+C_{\rm J}\frac{dV}{dt}  
-\sqrt{\frac{2k_{\rm B}T}{R_{\rm S}h_t}}&f_{\rm RND}(t).
\label{Eq1} 
\end{align}
The first equation corresponds to the voltage drop through the circuit while the second one accounts for the total current. The first term in the second equation is the current bias $I_{\rm b}$ and the second one is the feedback term. This term compares the voltage at two different times, $t$ and $t-t_{\rm D}$, and adds to the circuit, multiplied by a coupling constant $k$ that has units of inverse resistance. Since the feedback element may not allow for a full resolution of $\tau_{\rm J}$, we introduce an average for time-scales $\Delta t>\tau_{\rm J}$. The third term gives the current through the Josephson junction, with $I_{\rm c}$ the critical current and $\varphi$ the phase difference. The fourth and fifth terms provide the currents through the resistance $R_{\rm S}$ and the capacitor $C_{\rm J}$, respectively. The last term simulates the current fluctuations in the junction as a result of thermal noise, with $T$ the temperature and $f_{\rm RND}(t)$ a time-dependent function that provides normally distributed random values that average to zero with standard deviation of one.  This term is modeled following Ref.~\cite{noise_RK}, which offers an expression especially suitable for Runge-Kutta solvers ($h_t$ is the time discretization). 

We can now rewrite the set of first-order differential equations of Eqs.~\eqref{Eq1b} and~\eqref{Eq1} as a (dimensionless) second-order differential equation 
\begin{multline} 
\frac{d^2\varphi}{d\tau^2}=\frac{i_{\rm b}-\sin(\varphi)-\frac{d \varphi}{d \tau}}{\beta} \\ -\frac{kR_{\rm S}}{\beta}\frac{1}{\Delta \tau} \int_{\tau}^{\tau+\Delta \tau}\left( \frac{d\varphi}{d \tau}(\tau'-\tau_{\rm D})-\frac{d\varphi}{d \tau}(\tau')  \right)d\tau' \\ +\frac{1}{\beta}\sqrt{\frac{2\pi}{I_{\rm c}\Phi_0}}\sqrt{\frac{2k_{\rm B}T}{h_\tau}}f_{\rm RND}(\tau),
\label{Eq2} 
\end{multline}
where $\tau= t/\tau_{\rm J}$ is the reduced time (with $\tau_{\rm J}=\frac{\Phi_0}{2\pi I_{\rm c}R_{\rm S}}$ the inverse of the characteristic Josephson frequency), $\beta=\frac{2\pi I_{\rm c}R_{\rm S}^2 C_{\rm J}}{\Phi_0}$ the McCumber parameter, and $i_{\rm b}\equiv I_{\rm b}/I_{\rm c}$ the reduced bias current. Notice that $\tau_{\rm D}=t_{\rm D}/\tau_{\rm J}$ and $\Delta\tau=\Delta t/\tau_{\rm J}$ are also written in units of $\tau_{\rm J}$.

Defining $x_1=\varphi$ and $x_2=\frac{d\varphi}{dt}$, we can write
\begin{multline}
\frac{dx_2}{d\tau}= \frac{i_{\rm b}-\sin(x_1)-x_2}{\beta}  \\ 
-\frac{kR_{\rm S}}{\beta}\frac{1}{\Delta\tau}\int_{\tau}^{\tau+\Delta\tau}\left[x_2(t'-\tau_{\rm D})-x_2(t')\right]dt' \\ +\frac{1}{\beta}\sqrt{\frac{2\pi}{I_{\rm c}\Phi_0}}\sqrt{\frac{2k_{\rm B}T}{h_\tau}}f_{\rm RND}(\tau).
\label{Eq3}
\end{multline}
This equation is equivalent to Eq.~\eqref{RCSJFeed} of the main text but including the effect of a finite temperature through a white noise term. Moreover, we assume in the main text perfect coupling between the feedback element and the circuit, provided by $kR_{\rm S}=0.5$, in which the feedback perfectly compensates the voltage drop through the circuit (at $\Delta\tau\rightarrow0$). 

We solve this non-linear delayed differential equation using a fourth-order Runge-Kutta method as a function of time and sweeping for different bias current $i_{\rm b}$. For each current bias $i_{\rm b}$, we impose as initial condition the values of $(x_1,x_2)$ found for the previous value of $i_{\rm b}$. In this way we obtain a hysteretic behavior as also found in the experiments. We employ a small time discretization step $h_\tau$ so that the time scale $\tau_{\rm J}$ is well-resolved ($h_\tau=0.1\tau_{\rm J}$) and we compute the solution for very large time scales so that one can account for the phenomena happening at scales of $\tau_{\rm D}$. Notice that there are five orders of magnitude between $\tau_{\rm J}$ and $\tau_{\rm D}$, which makes these simulations computationally challenging.

Once $x_1(\tau)$ and $x_2(\tau)$ are found we compute the current and voltage through the Josephson junction as
\begin{equation}
    I(\tau)=I_{\rm c}\sin\left(x_1(\tau)\right), \; \; \; V(\tau)=R_{\rm S}I_{\rm c} x_2(\tau),
\end{equation}
and find their time average $\left<I\right>$ and $\left<V\right>$. We remove the initial solutions to the differential equation in the time average (a few $\tau_{\rm D}$ intervals) as the simulations may take some time to reach a possible different regime when changing $i_{\rm b}$. We also run several times (10 times) the same simulation using the same $i_{\rm b}$ and initial condition, and average among the solutions to remove the dependence on a particular realization of thermal noise.

\section{Feedback driven RCSJ model and experiment as a function of temperature}

We show the dependence of the current-voltage $\left<I\right>-\left<V\right>$ characteristic for different temperatures in Supplementary Fig.\,\ref{Josephson_T_R}. We observe that temperature fluctuations suppress the value of the current at the peak (the peak reaches $\sim 50$~nA at $T\rightarrow 0$). To a lesser degree, the extension of the bistable regime is also decreased with temperature and may actually disappear for too high temperatures. On the contrary, the time-averaged current $\left<I\right>$ in the bistable regime is barely affected by temperature. To understand this behavior, we note that at $T\rightarrow 0$ the entire stable regime happens just at $\left<V\right>=0$, while the bistable regime happens for a wide range of $\left<V\right>$. Temperature mixes the regimes close to their transitions, thus suppressing the peak at $\left<V\right>\rightarrow 0$ (and moving it to larger $\left<V\right>$ values) as it hybridizes with the bistable regime. On the other hand, $\left<I\right>$ in the central part of the bistable regime is barely affected by temperature, as it hybridizes with other $\left<V\right>$ values inside the same bistable regime. This also implies that whenever the thermal energy is comparable to or larger than the voltage range of the bistable regime, no bistability is possible. Analytically, we find that the time-averaged current within the bistable regime, just prior to transitioning to the AC Josephson regime, is roughly given by 
\begin{equation}
    \left<I\right>_{\rm bistable}\simeq I_{\rm c}\left(\frac{1-2kR_{\rm S}}{1-kR_{\rm S}}+\frac{\Delta\tau}{2\tau_{\rm D}}\frac{\left(kR_{\rm S}\right)^2}{1-kR_{\rm S}}\right).
    \label{Eq:I_bistable}
\end{equation}
We have checked that this analytical expression agrees with our numerical data either when $\Delta\tau$ is negligible, and thus the first term dominates, or when $\Delta\tau$ is comparable to $\tau_{\rm D}$ so that the second one does. For intermediate values, the behaviour of the system is more complex than the one this analytical equation can predict. Notably, this equation aligns well with the experimental data, as in our experimental setup $\Delta\tau$ is comparable to $\tau_{\rm D}$. Particularly, assuming perfect coupling $kR_{\rm S}=0.5$ and $\Delta\tau \simeq0.05\tau_{\rm D}$, this equation allows us to estimate a critical current $I_{\rm c}$ of 50~nA for the junction discussed in the Fig.\,\ref{Josephson}a of the main text, in agreement with Ambegaokar-Baratoff theory. Further support for this expression is demonstrated in Fig.~\ref{Fig_OscDepen} of the main text, where the current within the bistable regime obtained through experimental measurements closely follows the theoretical prediction of Ref.\,\cite{PhysRevB.51.3743} across a wide range of tunneling conductances.

\section{Parameters of the feedback-driven RCSJ model}

To explore the impact of different parameters of the feedback on the results of the model we calculate the time-averaged current $\left<I\right>$ for different bias currents $i_{\rm b}$ and parameters: $kR_{\rm S}$ (shown in Supplementary Fig.\,\ref{Parameters}a, $\left<I\right>$ is given by the color scale), $\Delta\tau$ (Supplementary Fig.\,\ref{Parameters}b), $\beta$ (Supplementary Fig.\,\ref{Parameters}c) and $\tau_{\rm D}$ (Supplementary Fig.\,\ref{Parameters}d) at zero temperature. In the stable regime, $i_{\rm b}\leq 1$ (Supplementary Fig.\,\ref{Parameters}a), the coupling constant has no significant influence since the feedback does not play a significant role. However, for $i_{\rm b}>1$ we observe  a substantial bistable regime that appears within the range of approximately $kR_{\rm S}\simeq 0.4$ to $kR_{\rm S}=0.5$, as indicated by the red area in Supplementary Fig.\,\ref{Parameters}a (red indicates a finite $\left<I\right>$, see color scale on the right). The saddle points of the differential equation determine the bistablility regime, which follows in this case $i_{\rm b}^{\rm (max)}\propto\left(\frac{1-kR_{\rm S}}{1-2kR_{\rm S}}\right)$. There is a divergence at $kR_{\rm S}=0.5$,  which serves as the upper boundary for the bistablility range. Conversely, the  lower boundary for $kR_{\rm S}$ is when $i_{\rm b}^{\rm (max)}=1$. Below this value the range for the bistable regime is infinitesimally close to the stable regime. In the present simulation this occurs at approximately $kR_{\rm S}\simeq0.4$. 

The effect of the integration interval $\Delta\tau$ is very similar to that of $kR_{\rm S}$ as it ultimately modifies the coupling between $x_2(\tau)$ and an average near $x_2(\tau-\tau_{\rm D})$. Thus, even at perfect coupling $kR_{\rm S}=0.5$, see Supplementary Fig.\,\ref{Parameters}b, the bistability is possible when $\Delta\tau$ is smaller than the characteristic delay time $\tau_{\rm D}$, as $x_2(\tau-\tau_{\rm D})$ is averaged to a very different quantity. Notice, nevertheless, that the bistable regime disappears when $\Delta\tau\sim \tau_{\rm D}$, as the feedback element averages to a time-independent constant.

The relationship between the bistable regime and the McCumber constant $\beta$ (Supplementary Fig.\,\ref{Parameters}c) is more intricate. In principle, one might argue that the saddle points of the differential equation are independent of $\beta$, suggesting that the bistablility range should remain unaffected by it. However, the value of $\beta$ does influence the rate at which the differential equation converges to the saddle points, which in turn influences the extension in parameter space of the bistable regime. 

To better understand this, let us explain in more detail the evolution of the system with time (for $\Delta\tau\rightarrow 0$). As discussed in the main text, the system undergoes a periodic evolution (with period $2\tau_{\rm D}$) in the bistable regime, oscillating between two distinct saddle points. Initially, the system reaches a periodic fixed point that persists for a time $\tau_{\rm D}$. This fixed point can be linked to an AC Josephson regime but with an average voltage that is renormalized to $\left<x_2\right>=i_{\rm b}/(1-kR_{\rm S})$ instead of the conventional $i_{\rm b}$. After this time $\tau_{\rm D}$, the action of the feedback element changes, leading the system to transition towards a different fixed point for an additional period $\tau_{\rm D}$. This latter saddle point is stable and exhibits similarities to a DC Josephson regime albeit with a renormalized current given by $x_{1}=i_{\rm b}\left(\frac{1- kR_{\rm S}}{1-2kR_{\rm S}}\right)$. Notably, the transition between these two saddle points occurs continuously in time. The system passes through an intermediary saddle point during this evolution. This intermediate point corresponds to the conventional AC Josephson regime (i.e., with $\left<x_2\right>=i_{\rm b}$), but it only serves as a fixed point for a (brief) period of time when the condition $x_2(t-\tau_{\rm D})\simeq x_2(t)$ is (approximately) satisfied. As the bias current $i_{\rm b}$ is increased the transition between the two primary fixed points is slower until reaching a critical value $i_{\rm b}^{\rm (max)}$ beyond which the evolution becomes so sluggish that the system remains solely in the conventional AC regime.

The value $i_{\rm b}^{\rm (max)}$ is also influenced by $k$, $\Delta\tau$ and $\beta$ since the rate of the transition is dependent on these parameters as well. In fact, one can analytically show that the eigenvalues of the Jacobian of the linearized differential equations near the saddle points exhibit a dependence on $\beta$, and specifically the real part scales as $\sim1/\beta$. This implies an exponential decay of the eigenstates towards the saddle points with a rate of $\sim1/\beta$, and so does $i_{\rm b}^{\rm (max)}$. Our numerical simulations, see Supplementary Fig.\,\ref{Parameters}c, confirm this intuition by showing that the extent of the bistable regime (red area) decreases approximately like $\sim1/\beta$. Note however that $\beta=0$ is a especial case in which there is no bistable regime.

We finally explore the dependence of the bistable regime with the delay time $\tau_{\rm D}$ in Supplementary Fig.\,\ref{Parameters}d. We observe that the value of $\tau_{\rm D}$ does not influence the bistable regime except when $\tau_{\rm D}$ is comparable to $\tau_{\rm J}$, i.e., $\tau_{\rm D}\lesssim 10^2$, where the behavior is more complex. This is because the evolution between the stable points becomes slower than $\tau_{\rm D}$ itself, thereby disrupting the bistablility regime.

\section{Measurement of the time dependent signal in the bistable regime with STM}

To obtain spatial maps in the bistable range, we have used a lock-in amplifier at the frequency of the bistable modulation signal (7.4 kHz here). We measure the output of the current-voltage converter with the lock-in without any further amplification, and obtain the magnitude of the oscillatory signal. As seen in Fig.\,\ref{Josephson}b,d, the voltage oscillates spontaneously in the bistable regime (i.e. without the action of an external drive). This leads to a current oscillation read by the current-voltage converter (Supplementary Fig.\,\ref{Scheme}). The amplitude of the oscillation is proportional to the magnitude of the range in bias where the bistable regime appears (Fig.\,\ref{Josephson}a,c). To make spatial maps of the Josephson critical current, it is particularly convenient to just trace the amplitude of the oscillatory signal using a lock-in amplifier measuring at the frequency of the oscillation, $2\pi/\tau_D$, as a function of the position at atomic scale.

We show in Supplementary Fig.\ref{LockIn}a the tunneling conductance as a function of the time-averaged voltage for different tunneling conductances, using tip and sample of Pb. We observe the peak at zero bias, due to the Josephson effect. In previous work (Refs.~\cite{Hamidian2016, doi:10.1126/science.abd4607, Cho2019, PhysRevLett.119.147702, Huang2020, Trahms2023, PhysRevB.64.212506, PhysRevLett.87.097004,PhysRevB.80.144506,PhysRevB.93.161115,Rodrigo04,10.1063/1.122654,Rodrigo2006PhysC,GUILLAMON2008537,PhysRevB.93.020504,KOHEN200518,PhysRevB.78.140507,Edkins976,Ast2016,Joo2019,doi:10.1073/pnas.2206481119,Liu2021,PhysRevLett.130.177002}) this peak is used to follow the magnitude of the (strongly smeared by thermal fluctuations) Josephson current as a function of the position and obtain atomic scale Josephson current maps. Note the presence of two negative resistance peaks at finite bias. The peak at low bias occurs on entering the bistable regime and the second one at larger bias when leaving the bistable regime. In Supplementary Fig.\ref{LockIn}b we show the finite frequency signal simultaneously observed in the lock-in amplifier as a function of the time-averaged voltage. We clearly see that the time dependent modulation only occurs in the bistable range and that its magnitude is large, making it easy to follow as a function of the position. The peak in the bistable regime is due to the time dependent oscillation shown in Fig.\,\ref{Josephson}b, and the oscillatory signal is zero elsewhere. The amplitude of this peak is traced as a function of the position to make the maps shown in Supplementary Figs.\,\ref{PDW},\ref{SwitchMap}.

We find that the spontaneous oscillatory signal obtained after the amplifier is as large as a fraction of a V (Supplementary Fig.\ref{LockIn}b). We can estimate the signal to noise ratio as the current at zero bias minus the current at large bias in the usual Scanning Josephson microscope (Supplementary Fig.\ref{LockIn}a), and compare it with the signal to noise ratio of the feedback driven atomic scale Josephson microscope, which is the oscillatory signal at the peak minus the signal outside the bistable range. We find that the signal to noise ratio is at least an order of magnitude larger in the Feedback driven atomic scale Josephson microscope. This also allowed us to observe the PDW in a small field of view, as shown in Supplementary Figure 6.

\section*{Data availability.} \
\noindent The data generated in this study have been deposited in the osf.io database under accession code https://doi.org/10.17605/OSF.IO/3KCD6\,\cite{Data}.

\section*{Acknowledgments.} \
\noindent Authors particularly acknowledge Juan Antonio Higuera from SEGAINVEX at UAM for the insight into the amplification circuitry used in STM. We also acknowledge discussions with Sebasti\'an Vieira about SJS. Support by the Spanish Research State Agency (PID\-2020-114071RB-I00, PID\-2020-117671GB-100, PDC2021-121086-I00, TED2021-130546B\-I00, PID2023-150148OB-I00 and CEX2018-000805-M), the European Research Council PNICTEYES through grant agreement 679080, the EU through grant agreement No 871106, grants No. PID2021-122769NB-I00 and No. PID2021-125343NB-I00 funded by MCIN/AEI/10.13039/501100011033, and by the Comunidad de Madrid through program NA\-NOFRONTMAG-CM (S2013/MIT-2850) is acknowledged. We have benefitted from collaborations through EU program Cost CA21144 (superqumap), and from SEGAINVEX at UAM in the design and construction of STM and cryogenic equipment.

\section*{Author contributions.} \
\noindent Calculations were carried out by S.D.E. and D.P., being supervised by E.P. and A.L.Y. The oscillatory signal was observed by J.A.M. and V.B. with the supervision of I.G. and H.S.. Further experiments were carried out by M.F.L., M.A. and B.W., with the supervision of E.H., H.S. and I.G. The electronics, data and set-up were analyzed by J.A.M., E.H. and J.G.R. The concept was devised by I.G., H.S. and A.L.Y. The manuscript was written by S.D.E., D.P., J.A.M., A.L.Y. and H.S. All authors have read and approved the final manuscript.

\section*{Ethics Declaration} \
\noindent The authors declare no competing interests.

\clearpage

\clearpage

\appendix

\section*{Appendix A: Temperature dependence of the bistable regime and results in Al and in 2H-NbSe$_2$}
\label{secTempAlNbSe2}

We show data taken with a tip of Pb ($T_{\rm c}^{\rm (Pb)}=7.2$~K) and a sample of 2H-NbSe$_2$ ($T_{\rm c}^{\rm (NbSe_2)}=7.15$~K) in Extended Data Fig.\,3. 2H-NbSe$_2$ is known to exhibit multiple superconducting gaps $\Delta_{\mathrm{2H-NbSe_2},i}$, which are distributed over the Fermi surface, ranging from $\Delta_{\mathrm{2H-NbSe_2},1}\approx0.5$~meV to $\Delta_{\mathrm{2H-NbSe_2},2}\approx1.5$~meV~\cite{Guillamon2008, Liu2021, Sanna2022, doi:10.1126/science.1065068, PhysRevB.92.134510, PhysRevB.76.212504, PhysRevLett.90.117003, RODRIGO2004306}. This leads to the rounded current-voltage characteristics above about $1$~mV shown in the upper inset of Extended Data Fig.\,3a. Despite these differences, the behavior regarding the bistable regime is similar to Pb-Pb junctions. 

In Extended Data Fig\,.4, we show the results for the Al-Al junction ($T_{\rm c}^{\rm (Al)}=1.2$~K). Due to the reduced superconducting gap compared to Pb, the voltage range is modified. Furthermore, the peak in the Josephson current is considerably reduced and the hysteresis is practically absent (Extended Data Fig.\,4). These curves were also taken at a relatively large temperature of 0.4~K. Nonetheless, the time-dependent oscillatory behavior in the bistable regime is still observed (Extended Data Fig.\,4b). We employed for these measurements a capacitor with ten times higher capacitance in the operational amplifier circuit than in the other experiments. This increased the time constant $\tau_{\rm D}$ by about an order of magnitude.

\section*{Appendix B: Entry and exit of the bistable regime}
\label{sec3}

We can analyze in more detail the effect of temperature by studying the entry into and exit out of the bistable regime.

The escape dynamics of small Josephson junctions has been extensively studied~\cite{PhysRevLett.77.3435, PhysRevLett.55.1908, doi:10.1063/1.3699625, Tafuri19, PhysRevLett.109.050601, Joyez1999}. The current jumps from the Josephson regime into the resistive regime with its dynamics determined by thermal activation at high temperatures and quantum tunneling at very low temperatures~\cite{PhysRevLett.77.3435, PhysRevLett.55.1908}. This behavior is radically different from what we observe. Whereas the escape dynamics is determined by the junction jumping between a state with Cooper pair tunneling to a state without Cooper pair tunneling, the dynamics of the bistable regime is between two different Cooper pair tunneling states.

In the bistable regime, the junction leaves the zero voltage regime and enters the bistable regime, but only Cooper pairs are exchanged between both sides of the junction. In Extended Data Fig\,.7a, we show that this occurs through a stochastic process. Pulses of width $\tau_D$ are measured at time intervals that seem to be random. The inverse of the time between pulses, $\frac{1}{\tau_{\rm Sto}}$, increases exponentially with bias. We can find the probability $P_{1}(V)$ by taking $P_{1}(V)=\tau_{\rm Sto}^{-1}\left(\int_0^1 P(t)dt\right)$. $P_{1}(V)$ follows a Gaussian behavior within a range of time-averaged voltage of about $5$~$\mu V$ (see Extended Data Fig.\,7b).

The junction leaves the bistable regime producing the hysteresis discussed above. To analyze this behavior, we have swept through the current-voltage characteristics many times and traced at each sweep the voltage where we lose oscillations at $\tau_{\rm D}$. From this, we define a probability $P_{2}$ and the corresponding time scale (Extended Data Fig.\,7a). As we observe in Extended Data Fig.\,7b, the Gaussian function that describes the  probability becomes sharper when decreasing temperature.

\section*{Appendix C: Other approaches to measure the Josephson current}
\label{OtherJosephson}

The current vs voltage characteristic in SJS is often understood through the phase diffusive transport approximation derived by Ivanchenko and Zilberman~\cite{IZ1969}. According to this model, the current is given by $I=\frac{I_{\rm c}^2}{2}Z\left(\frac{V}{V^2+V_{\rm c}^2}\right)$, where $Z$ represents a high-frequency impedance and $V_{\rm c}=\frac{2e}{\hbar}k_{\rm B}TZ$ is a characteristic voltage. The conductance at zero bias $\frac{dI}{dV}(V=0)$ is often used to map the Josephson current as a function of the position. However, $Z$ is typically unknown~\cite{PhysRevLett.87.097004}. Furthermore, the maximal current (for $V=V_c$) decreases as $1/T$ and the conductance at zero bias as $\frac{dI}{dV}(V=0)$ as $1/T^2$.

In the Fig.\,1c of the main text we compare curves taken in different STM set-ups (blue line and dashed blue line) using tip and sample of Pb. We have tried to match as far as possible the conductance $G$, which is around $0.4G_0$. In the set-up having a capacitance of around $C\approx 0.3$~nF (blue curve) we observe the bistable regime. In the set-up having a capacitance of about $C\approx 10$~nF (dashed blue curve) we observe no bistable regime. The current and voltage are then time-independent and there is no hysteresis. The dashed blue curve is well captured by the Ivanchenko-Zilberman approximation.

Our model also requires an effective temperature of 1~K to obtain the shape of the curve at low bias voltages (Fig.~3c in the main text). It is important to note that the energy resolution of our experiment is far below this value, as evidenced by the sharp increase in the current at the gap edge (inset of Fig.~1c, and Ref.~\cite{doi:10.1063/5.0059394}). The increased temperature arises from the equivalent temperature of the electromagnetic environment at the Josephson frequency $\omega_{\rm J}$. However, we can extract the value of $\left<I\right>$ in the bistable regime from the oscillatory signal, which is given by $I_{\rm c}$ and the time constants of our circuit, and is thus almost independent of temperature for $T<T_{\rm c}$/2, following the dependence of $I_{\rm c}$ with temperature.

\section*{Appendix D: Improved mapping of the pair density wave in 2H-NbSe$_2$ in the time-dependent bistable regime}
\label{PDW}

2H-NbSe$_2$ is a strongly anisotropic superconductor (T$_c=7.2$ K) presenting a charge density wave (CDW) below T$_{CDW}=33$ K\,\cite{Johannes2006}. The CDW is characterized by spatial modulations of the charge along the main crystalline directions of the hexagonal lattice, $\rho^i_{CDW}(\bm{r})=\rho_0 cos(\bm{Q^i_{CDW}}\bm{r})$. The absolute value of the CDW wavevector is a third of the lattice wavevector $\bm{Q^i}$, $\bm{Q^i_{CDW}} \approx \frac{1}{3}\bm{Q^i}$. In the Fourier transform of a topographic image (inset of the Extended Data Fig.\,6a), we can identify the six atomic lattice Bragg peaks (at $\bm{Q^i}$) and the peaks of the CDW (at $\bm{Q^i_{CDW}}$). The superconducting gap presents small spatial modulations at the atomic and CDW wavevectors\,\cite{Guillamon2008c,doi:10.1126/science.abd4607}. It was shown recently that small gap modulations occur at the same time as modulations in the amplitude of the Josephson current, demonstrating the occurence of a pair density wave (PDW) in 2H-NbSe$_2$\,\cite{doi:10.1126/science.abd4607}.

A PDW is a phase coherent state where the order parameter is spatially modulated\cite{doi:10.1146/annurev-conmatphys-031119-050711}. We can consider for simplicity a PDW state with an order parameter $\Delta_{PDW}$ varying as $\Delta_{PDW}(\bm{r})\propto cos(\bm{Q_{PDW}}\bm{r})$ (also called Larkin-Ovchinikov phase\,\cite{LO65,PhysRev.135.A550}). This state is time reversal symmetric but breaks translational symmetry, with an order parameter which is zero on spatial average. There are spatial fluctuations of the phase of the PDW, for example due to disorder, and we can write $\Delta_{PDW}(\bm{r})\propto cos(\bm{Q_{PDW}}\bm{r}+\theta(\bm{r}))e^{i\varphi(\bm{r})}$ where $\theta$ refers to the phase of the spatial PDW modulation with respect to the CDW modulation and $\varphi$ to the superconducting phase\cite{PhysRevX.5.031008}. The PDW of 2H-NbSe$_2$ occurs along the three CDW wavevectors $\bm{Q^i_{CDW}}$ and there are changes in $\theta$ at defects. Importantly, there is, in addition, an overall phase difference $\delta\theta=2\pi/3$ between PDW and CDW\,\cite{doi:10.1126/science.abd4607}. The overall phase difference has been associated to the difference in amplitude of the pair density along the CDW wavevectors with respect to the overall pair density, due to the anisotropy in the orbital character of the Cooper pair wavefunctions\,\cite{doi:10.1126/science.abd4607}. In Ref.\,\cite{doi:10.1126/science.abd4607}, the measurement of the Josephson current was made in the strongly diffusive regime, which is described using the  Ivanchenko and Zilberman approach. As discussed above, the conductance at zero bias $\frac{dI}{dV}(V=0)=\frac{I_{\rm c}^2}{2}Z\left(\frac{1}{V_{\rm c}^2}\right)$ is then proportional to $I_{\rm c}$ but strongly thermally smeared and temperature dependent. It seems natural to consider if the oscillatory behavior in the bistable regime leads to the same observations on the PDW of 2H-NbSe$_2$, with the advantage of providing a significant improvement of the signal to noise ratio, as we show in Extended Data Fig.\,10 and in Methods Section 5.

We show in Extended Data Fig.\,6 a simultaneous measurement of the topography (a), the feedback driven Josephson current (b) and the zero bias conductance (c). In the upper left insets we provide the respective Fourier transforms. We clearly observe the atomic lattice Bragg peaks and the CDW in the topography (Extended Data Fig.\,6 a). In the bistable regime (Extended Data Fig.\,6 b), the peaks at the CDW wavevectors are due to the PDW. The PDW is well resolved in the bistable regime (Extended Data Fig.\,6b), but is barely visible with the usual Josephson measurement (Extended Data Fig.\,6 c).

When Fourier filtering the images (at $\bm{Q^i_{CDW}}$) we find the maps shown in Extended Data Fig.\,6(d, e, f). We see that the PDW is shifted by $\delta\theta=2\pi/3$ with respect to the CDW as in Ref.\,\cite{doi:10.1126/science.abd4607}. Notice also that we could resolve the PDW in a much smaller field of view. This shows that the feedback driven Josephson microscope provides a significant improvement in the measurement of the Josephson coupling at atomic scale.

\section*{Appendix E: Pair density wave and current phase relation in 2H-NbSe$_2$ from the time-dependent bistable regime}
\label{sec5}

As we have shown above (Supplementary Information Section 2), the entry into and exit out of the bistable regime is thermally induced. In a conventional Josephson junction, it is known that switching and retrapping between Cooper pair tunneling and quasiparticle tunneling depends on the shape of the current-phase relation. For a perfectly symmetric current-phase relation with $I_c=sin(\varphi)$, as in a usual tunnel junction between s-wave superconductors, the switching and retrapping occurs at the same positions for negative and positive bias. But an asymmetric current-phase relation leads to a non-reciprocal behavior with respect to bias.

In Extended Data Fig.\,8a we show a simulation with a current-phase relation $I(\varphi)=sin(\varphi+\pi/2)+1/2sin(2\varphi)$. We see that the non-reciprocal behavior is also visible for the entry into and exit out of the bistable regime. Contrary to the usual switching and retrapping, in the bistable regime it occurs at a large finite time-averaged voltage.

This is not very surprising, as we are using similar equations as in the usual RCSJ model. However, the occurence at a larger time-averaged could be benefitial to explore the spatial dependence of the entry and exit behavior. To this end, we have plotted the exit point out of the bistable regime as a function of the position, giving the maps shown in Extended Data Fig.\,9. We observe again the PDW.

Deviations from the usual $I=I_c sin(\varphi)$ current phase relation have been predicted to occur in a PDW\,\cite{doi:10.1146/annurev-conmatphys-031119-050711}. In a most simple PDW state, the phase of the order parameter switches sign, suppressing the usual Josephson current and leading to higher order Josephson processes due to a modified current phase relation. Eventually, one could have that $I(\varphi)=I_1sin(\varphi)+I_2sin(2\varphi)$ with $I_1=0$\,\cite{doi:10.1146/annurev-conmatphys-031119-050711}. Josephson junctions consisting of arrays of spatially alternating $0$ and $\pi$ junctions present either spatially homogeneous or inhomogeneous coupling, depending on the size of the $0$ and $\pi$ regions and the penetration depth $\lambda$\,\cite{PhysRevB.67.220504}. In the PDW of 2H-NbSe$_2$ the size for phase changes is of the order of the CDW modulation, three times the interatomic distance and thus much smaller than the penetration depth of 2H-NbSe$_2$. Then, we can indeed expect spatially varying phase coupling. If the length of $0$ and $\pi$ regions is slightly different there is a rapidly varying phase change $\varphi_{PDW,r}(x)$ and a slowly varying  $\varphi_{PDW,s}(x)$\,\cite{PhysRevB.67.220504}. This leads to spatial variations in the current-phase relation $I(\varphi_{PDW,s}(x)+\varphi_{PDW,r}(x))$, consisting of the usual sinusoidal component and higher order terms\,\cite{doi:10.1146/annurev-conmatphys-031119-050711,PhysRevB.67.220504}. The spatial modulation obtained in our experiments (Extended Data Fig.\,9) suggests spatial variations in the current phase relation at the PDW wavevector, with, in addition, a slowly varying shift with respect to the PDW amplitude.

\section*{Appendix F: Relation to relaxation oscillators and other dynamic phenomena in Josephson junctions}
\label{sec6}

The nonlinear dynamics of a Josephson junction in itself, without feedback, can be extremely rich and complex, with regimes of chaotic behavior~\cite{PhysRevE.53.405, doi:10.1063/1.368113, PhysRevA.31.2509, PhysRevB.92.174532, PhysRevB.85.184302,Kleiner2021}. Additionally, unstable regions in the current-voltage characteristics of Josephson junctions can lock into passive elements leading to an oscillatory behavior~\cite{92493, PatentOscillator1, 621998, ONOMI2016141, doi:10.1063/1.321785, Muck1988, doi:10.1063/1.92782}. Often, they are called relaxation oscillators and imply a jump of the current-voltage characteristic to the dissipative regime with a voltage of order of the superconducting gap. The behavior discussed here occurs in a narrow interval close to zero, far below the superconducting gap value. Furthermore, relaxation oscillators and related behavior occur at frequencies between four and six orders of magnitude higher than those observed here~\cite{92493, PatentOscillator1, 621998, ONOMI2016141, doi:10.1063/1.321785, Muck1988, doi:10.1063/1.92782}.

On the other hand, Josephson junctions connected to a memory resistor (memristor) have been analyzed theoretically, finding different regimes that include oscillatory and chaotic behavior \cite{journal.pone.0191120,GRONBECHJENSEN1992131}. A memristor can be realized by an active circuit element \cite{Strukov2008,ChuaLO}. Our method to obtain a feedback is based on the amplifier as an active circuit element and inherently provides a system with memory (time-delayed input). The experimental realization of such a "memristor-like" system shown here provides an interesting playground to study the formation of chaos in nonlinear oscillators and its connection with networks and synchronization.


\clearpage 

\section*{Appendix G: Supplementary Figures}
\setcounter{figure}{0}

\captionsetup[figure]{labelfont={bf},name={Supplementary Figure \let\nobreakspace\relax},labelsep=none}

\begin{figure}
	\begin{center}
		\includegraphics[width=0.65\textwidth]{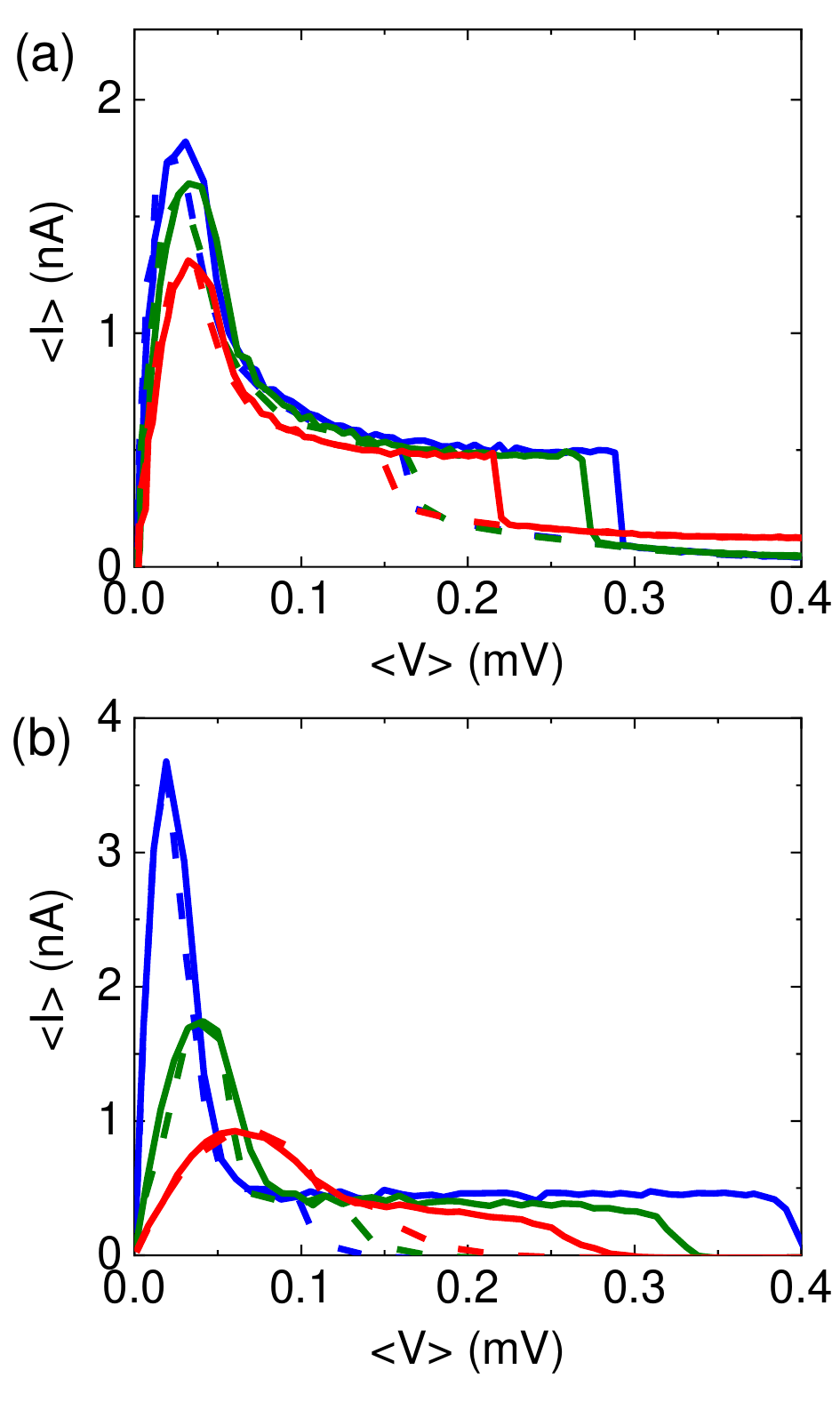}
	\end{center}
	\vskip -1cm
	\caption{\noindent {\bf \ |  Bistable regime at different temperatures.} {\bf a} Time-averaged current $\left<I\right>$ vs time-averaged voltage $\left<V\right>$ measured in a junction between a Pb tip and a Pb sample taken at different temperatures: $T=1$~K (blue), $T=2$~K (green) and $T=3$~K (red). To obtain the curve at 3~K we have subtracted the temperature-induced quasiparticle current. Solid lines are data taken with increasing bias while dashed lines are with opposite ramping direction. Tunneling conductance is $G\approx 0.1 G_0$, being $G_0$ the conductance quantum.  {\bf b} Results from the model calculations. We plot $\left<I\right>$ vs voltage $\left<V\right>$ at $T=0.5$~K (blue), $T=1.25$~K (green) and $T=2.5$~K (red).}
	\label{Josephson_T_R}
\end{figure}

\begin{figure}
	\begin{center}
		\includegraphics[width=0.99\textwidth]{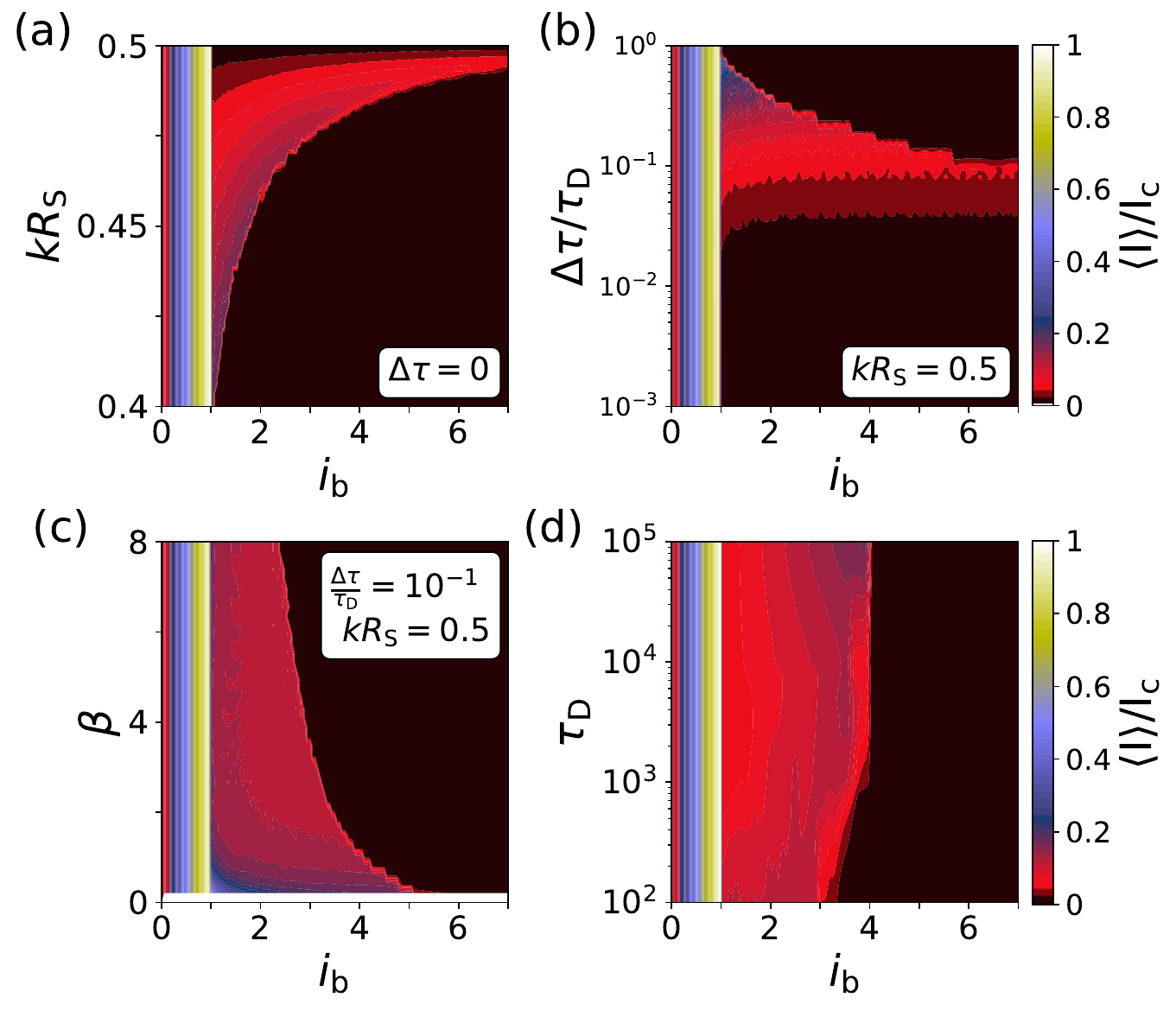}
	\end{center}
	\vskip -0.5cm
	\caption{\noindent {\bf \ | Evolution of the junction current for different parameters.} Evolution of the time-averaged current through the junction $\left<I\right>/I_{\rm c}$ as a function of the bias current $i_{\rm b}$ and: {\bf a} the feedback constant $k$ (while fixing $\Delta\tau=0$, $\tau_{\rm D}=10^{5}$ and $\beta=4$), {\bf b} the interval of integration $\Delta\tau$ ($kR_{\rm S}=0.5$, $\tau_{\rm D}=10^{5}$ and $\beta=4$),  {\bf c} the McCumber parameter $\beta$ ($kR_{\rm S}=0.5$, $\Delta\tau/\tau_{\rm D}=10^{-1}$ and $\beta=4$), and {\bf d} the delay time $\tau_{\rm D}$ ($kR_{\rm S}=0.48$, $\Delta\tau=0$ and $\beta=4$).}
	\label{Parameters}
\end{figure}

\begin{figure}
	\begin{center}
		\includegraphics[width=\textwidth]{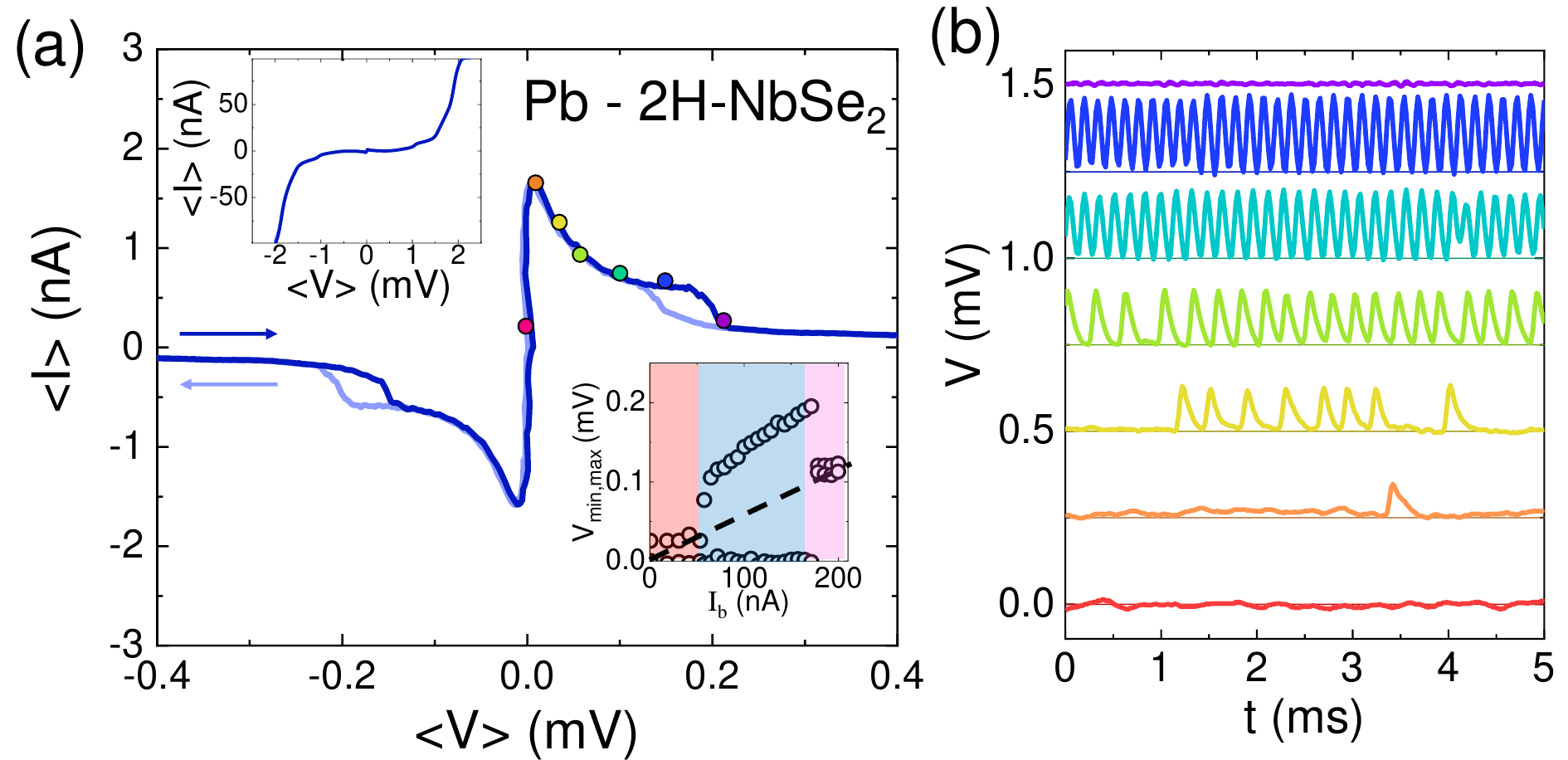}
	\end{center}
	\vskip -0.5cm
	\caption{\noindent {\bf \ | Feedback-driven Josephson effect in Pb-2H-NbSe$_2$.} {\bf a} Time averaged current $\left<I\right>$ vs the time-averaged voltage $\left<V\right>$ obtained with a tip of Pb and a sample of 2H-NbSe$_2$ is shown as lines. The colored dots represent the positions where we have measured the time-dependent voltage shown in {\bf b}.  In the lower inset we show the maximum and minimum voltages vs $I_b$ (see main text). {\bf b} The voltage as a function of time $V(t)$ is shown as colored lines. The $V(t)$ curves are vertically shifted for clarity by 0.25~mV. Temperature is 280~mK and the tunneling conductance is $0.4G_0$.}
	\label{PbNbSe2}
\end{figure}

\begin{figure}
	\begin{center}
		\includegraphics[width=\textwidth]{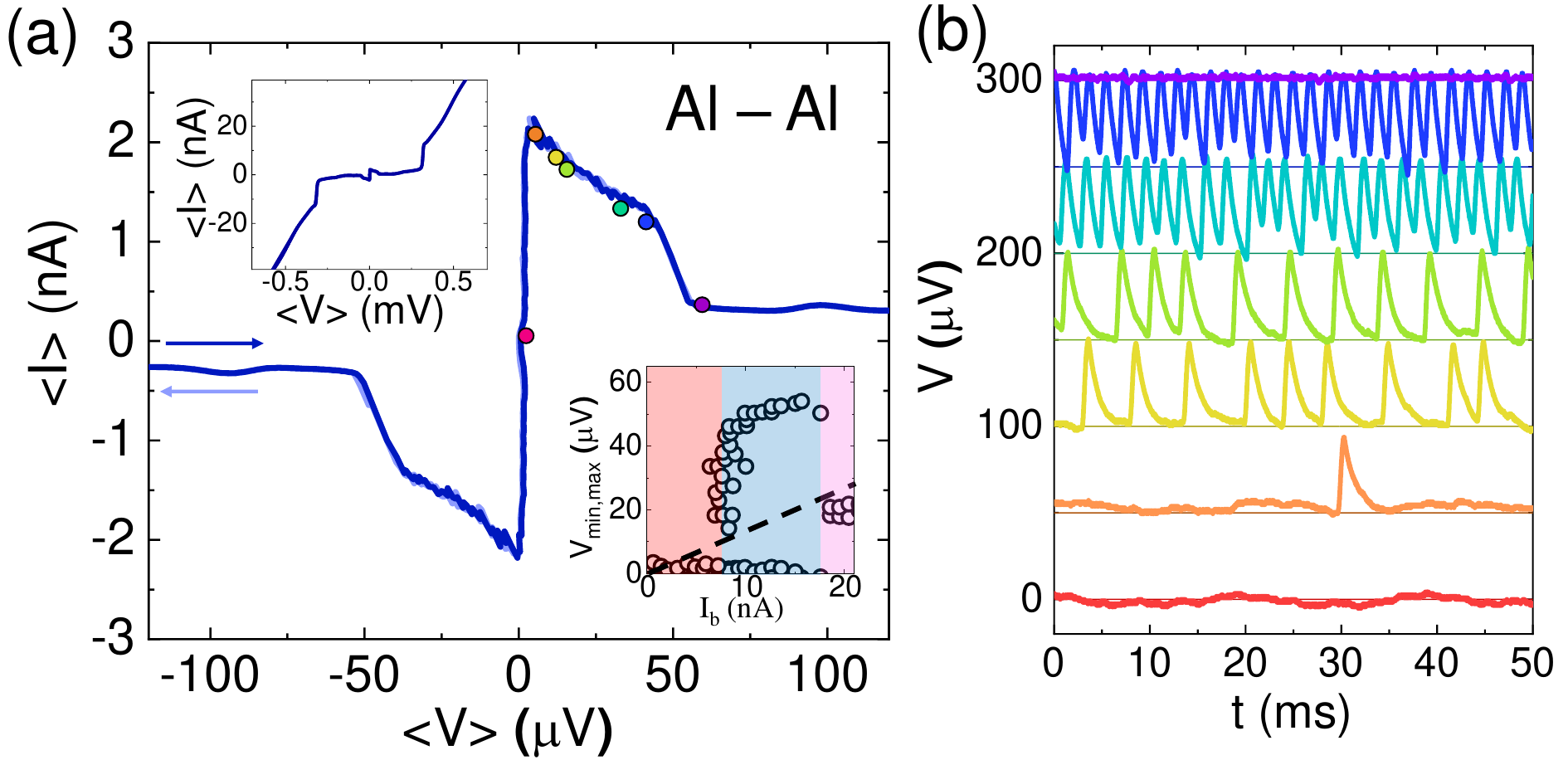}
	\end{center}
	\vskip -0.5cm
	\caption{\noindent {\bf \ | Feedback driven Josephson effect in Al-Al.} {\bf a} Time averaged current $\left<I\right>$ vs the time-averaged voltage $\left<V\right>$ obtained with tip and sample of Al is shown as lines. The colored dots represent the positions where we have measured the time-dependent voltage shown in {\bf b}. In the lower inset we show the maximum and minimum voltages vs $I_{\rm b}$. {\bf b} The voltage as a function of time $V(t)$ is shown as colored lines. The $V(t)$ curves are vertically shifted for clarity by 0.25~mV. Temperature is 400~mK and the tunneling conductance is $0.6G_0$.}
	\label{Al}
\end{figure}

\begin{figure}
	\begin{center}
		\includegraphics[width=\textwidth]{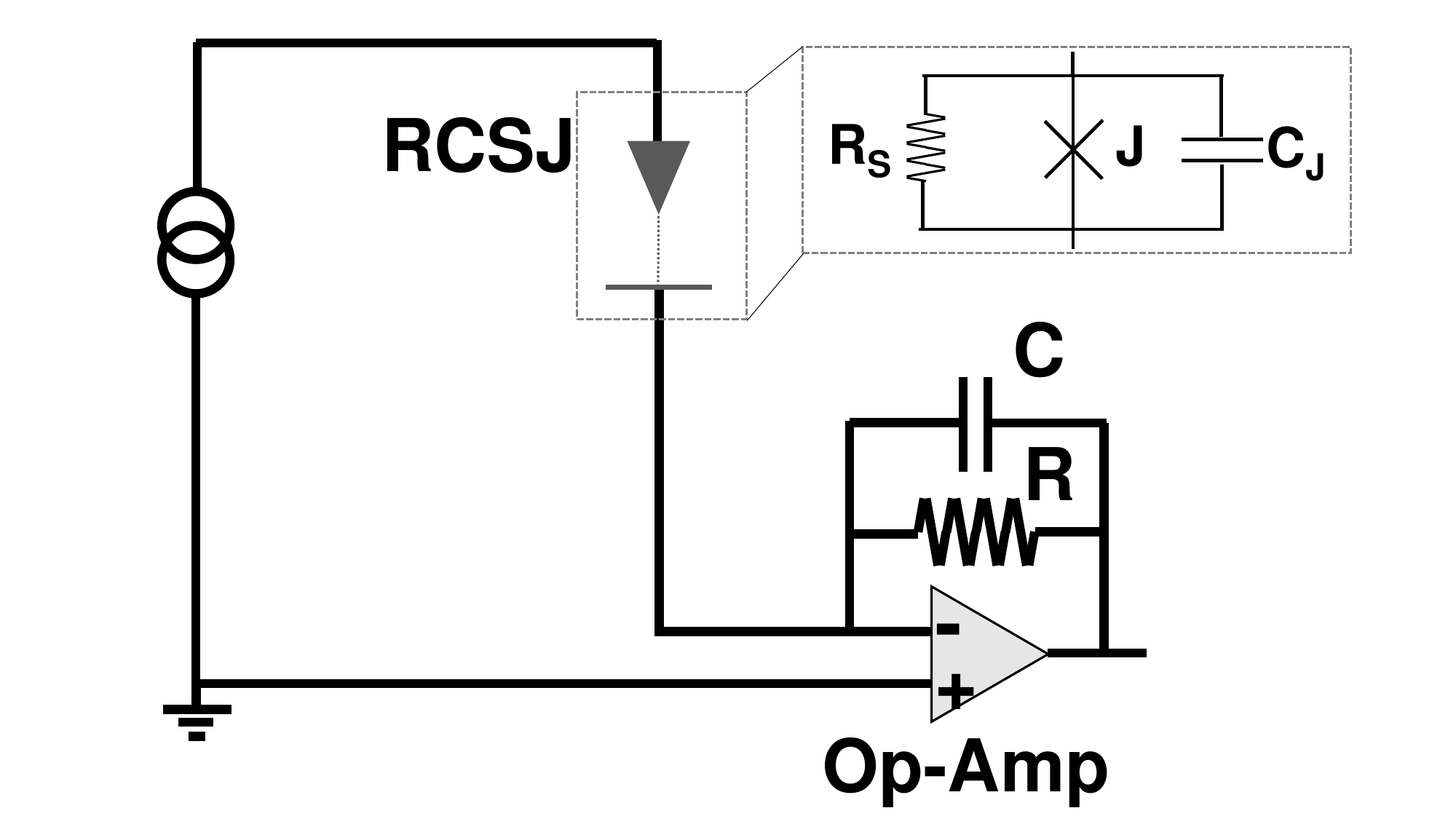}
	\end{center}
	\vskip -0.5cm
	\caption{\noindent {\bf \ | Scheme of a STM circuit.} Scheme of a typical measurement circuit used in STM. The junction is shown by a triangle (tip), the tunneling current (dashed line) and the sample (horizontal bar), enclosed in a dashed rectangle. The electromagnetic environment of the junction is modelled using a RCSJ model (upper right scheme in the dashed rectangle). The operational amplifier is shown schematically on the bottom right. The junction is driven by a source (circles) and the current through the junction is measured by the usual current-voltage converter circuit of the operational amplifier. The amplifier uses a feedback circuit whose time constant is given by the resistance and capacitance connected to the operational amplifier.}
	\label{Scheme}
\end{figure}

\begin{figure}
	\begin{center}
		\includegraphics[width=0.82\textwidth]{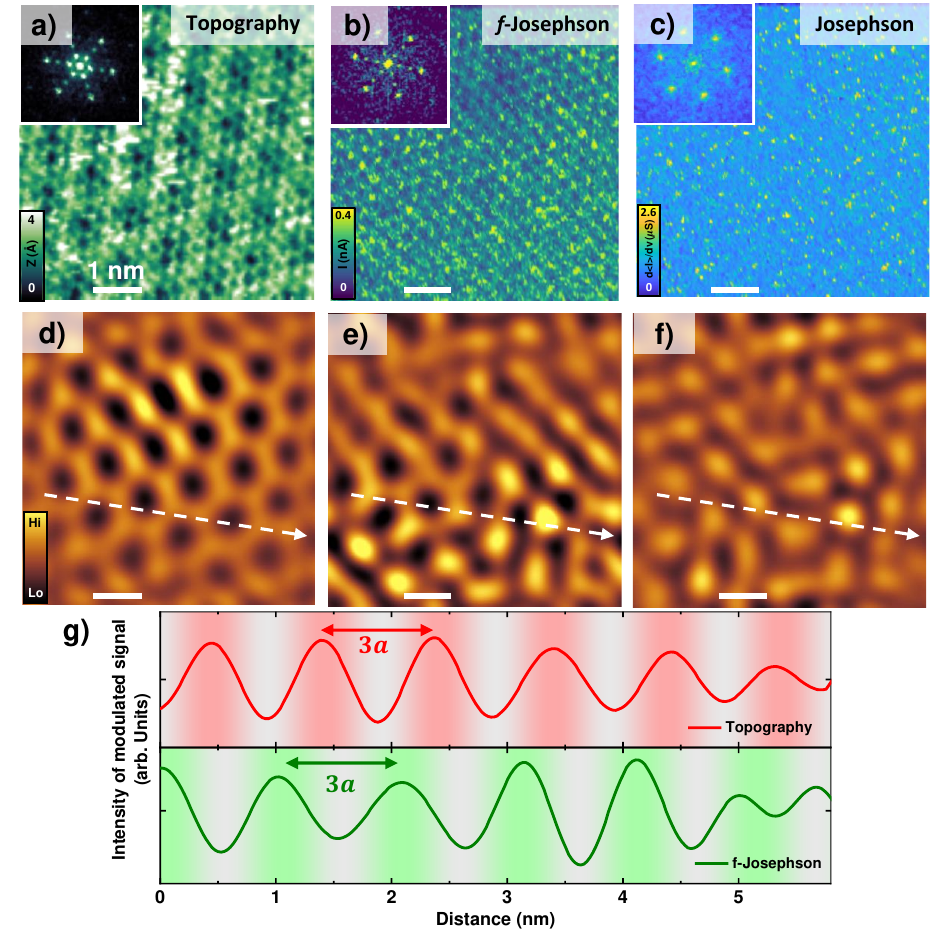}
	\end{center}
	\vskip -0.5cm
	\caption{\noindent {\bf \ | Pair density wave observed in the bistable regime.} {\bf a} STM topographic image of 2H-NbSe$_2$ taken with a tip of Pb ($T=$ 0.8 K and $G=0.17G_0$). {\bf b} Atomic scale map obtained by measuring the amplitude of the oscillatory signal in the bistable regime by directly reading the output of a lock-in amplifier. The amplitude of the oscillatory signal is proportional to the time-averaged voltage, as shown in Figs.\,3b,d of the main text. The amplitude is maximal at the exit point of the bistable range, which depends on $I_c$, as shown in Fig.\,4 of the main text. {\bf c} Atomic scale map of the zero bias tunneling conductance, obtained by tracing the current in the usual stable regime. Insets at the upper left show the Fourier transform. White scale bars are 1 nm long. {\bf d-f} Fourier filtered images at the PDW wavevectors. Color scale is adjusted for maximum contrast in each panel. Note that the color scale corresponds to a signal about three times smaller in {\bf f} than in {\bf e}, showing the important improvement obtained through the measurement of the current in the bistable regime. {\bf g} Linescan of the topography (red line, upper panel) and of the Josephson current in the bistable regime (green line, bottom panel, we take the oscillation along the direction of the white arrrows). We highlight schematically the CDW modulation in background red-grey in the upper panel and the PDW modulation in background green-grey in the bottom panel.}
	\label{PDW}
\end{figure}

\begin{figure}
	\begin{center}
		\includegraphics[width=\textwidth]{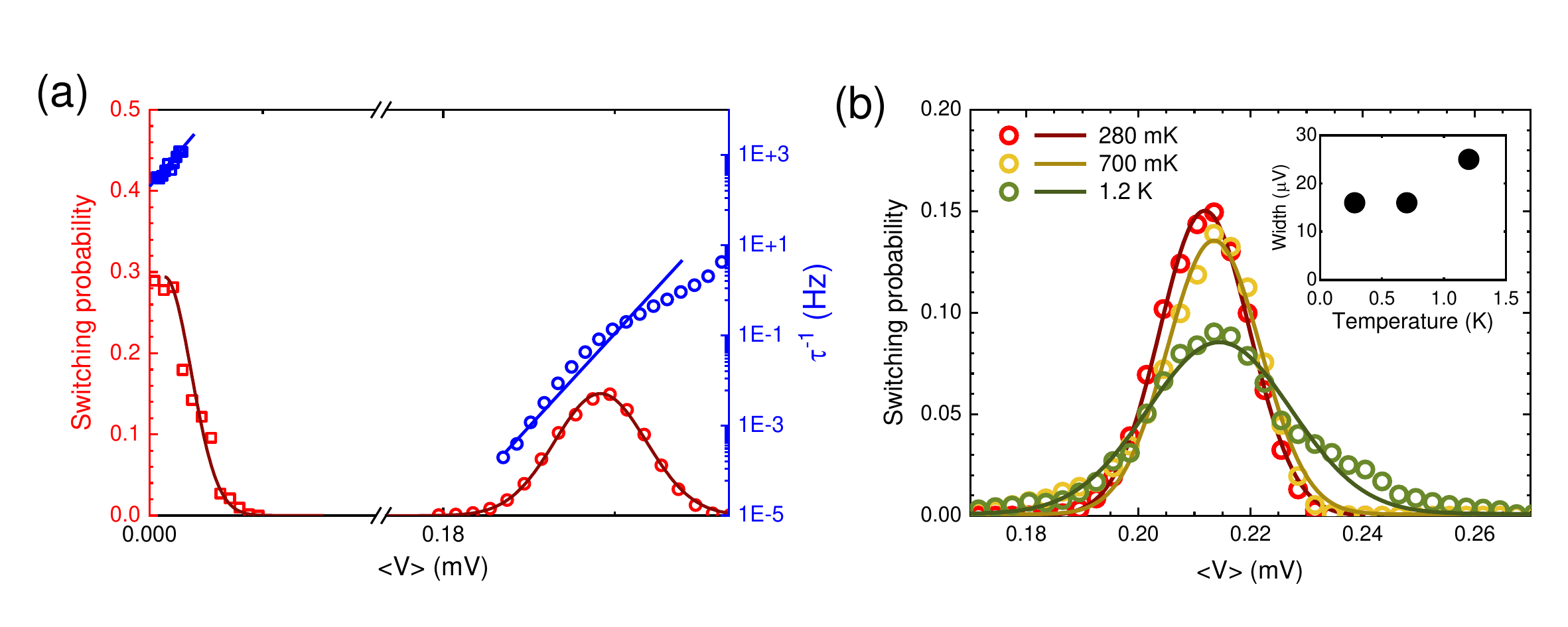}
	\end{center}
	\vskip -0.5cm
	\caption{\noindent {\bf \ | Entry and exit behavior of the bistable regime.} {\bf a} Probability (left $y$-axis) and time (right $y$-axis) are shown as a function of the time-averaged bias voltage $\left<V\right>$ at 280~mK. We show the entry into the bistable regime with squares and the exit out of the bistable regime with circles. Fits to a Gaussian (red points) and to an exponential (blue points) are shown as lines in grey and blue color. {\bf b} Probability (circles) vs time-averaged voltage $\left<V\right>$ is shown for different temperatures: $T=280$~mK (dark red), $T=700$~mK (orange), and $T=1.2$~K (green). Data fits with Gaussians are shown with solid lines with the same colors and the width (dispersion) of the Gaussians vs temperature is shown in the inset.}
	\label{Histo}
\end{figure}

\begin{figure}
	\begin{center}
		\includegraphics[width=\textwidth]{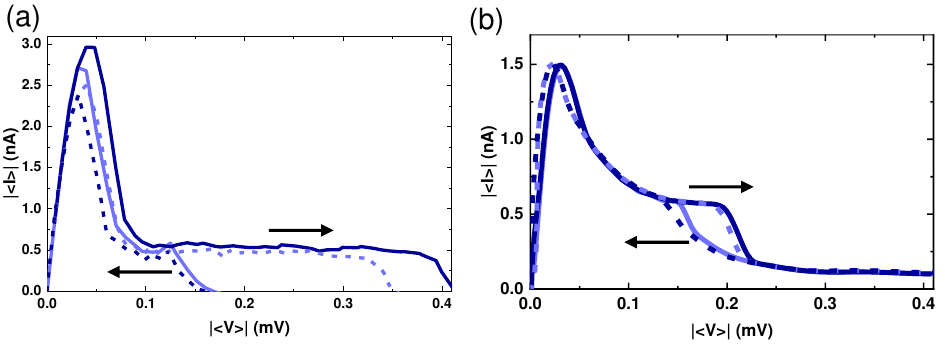}
	\end{center}
	\vskip -0.5cm
	\caption{\noindent {\bf \ | Anisotropy in the entry and exit behavior of the bistable regime.} {\bf a} The absolute values of the calculated time-averaged current vs time-averaged voltage curves, $\lvert \left<I\right>\rvert-\lvert \left<V\right>\rvert$, are shown as lines. We show as continous lines positive and as dashed lines negative values of $\left<I\right>$ and $\left<V\right>$. Black arrows indicate the direction of the variations in $\lvert \left<V\right>\rvert$ (increasing or decreasing). We see that the bistable regime (when $\left<I\right>$ is mostly independent of $\left<V\right>$), is left and entered at different values of $\lvert \left<V\right>\rvert$, depending on the direction of the variations in $\lvert \left<V\right>\rvert$ and on the polarity. {\bf b} Time-averaged current vs time-averaged voltage curves, $\lvert \left<I\right>\rvert-\lvert \left<V\right>\rvert$, obtained on 2H-NbSe$_2$ with a tip of Pb (average over $65000$ curves, $T=$ 0.28 K, $G=0.14G_0$).}
	\label{Switch}
\end{figure}

\begin{figure}
	\begin{center}
		\includegraphics[width=\textwidth]{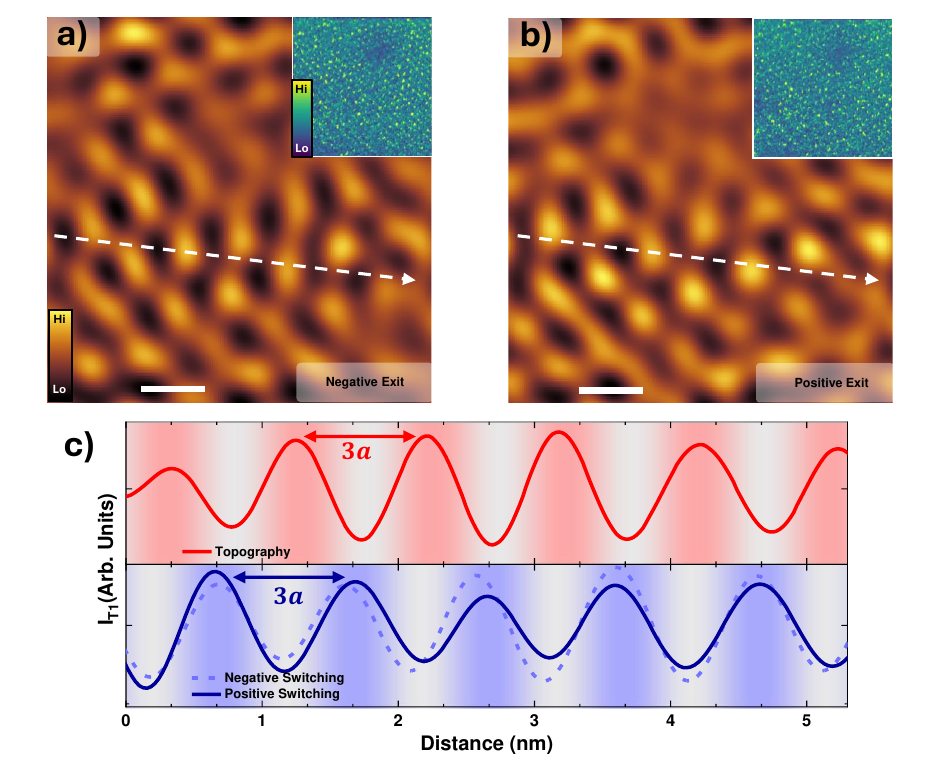}
	\end{center}
	\vskip -0.5cm
	\caption{\noindent {\bf \ | Map of the exit of the bistable regime.} {\bf a,b} In the main panels we show the spatial map of the exit of the bistable regime at negative ({\bf a}) and positive ({\bf b}) $\left<V\right>$. In the top right insets we show the unfiltered maps. White scale bars are 1 nm long. {\bf c} We show profiles along the white dashed lines in {\bf a,b} of the topography (red line, top panel) and the $\left<I\right>$ at the exit of the bistable regime (bottom panel, continuous dark blue line for positive $\left<V\right>$ and dashed light blue line for negative $\left<V\right>$).}
	\label{SwitchMap}
\end{figure}

\begin{figure}
	\begin{center}
		\includegraphics[width=\textwidth]{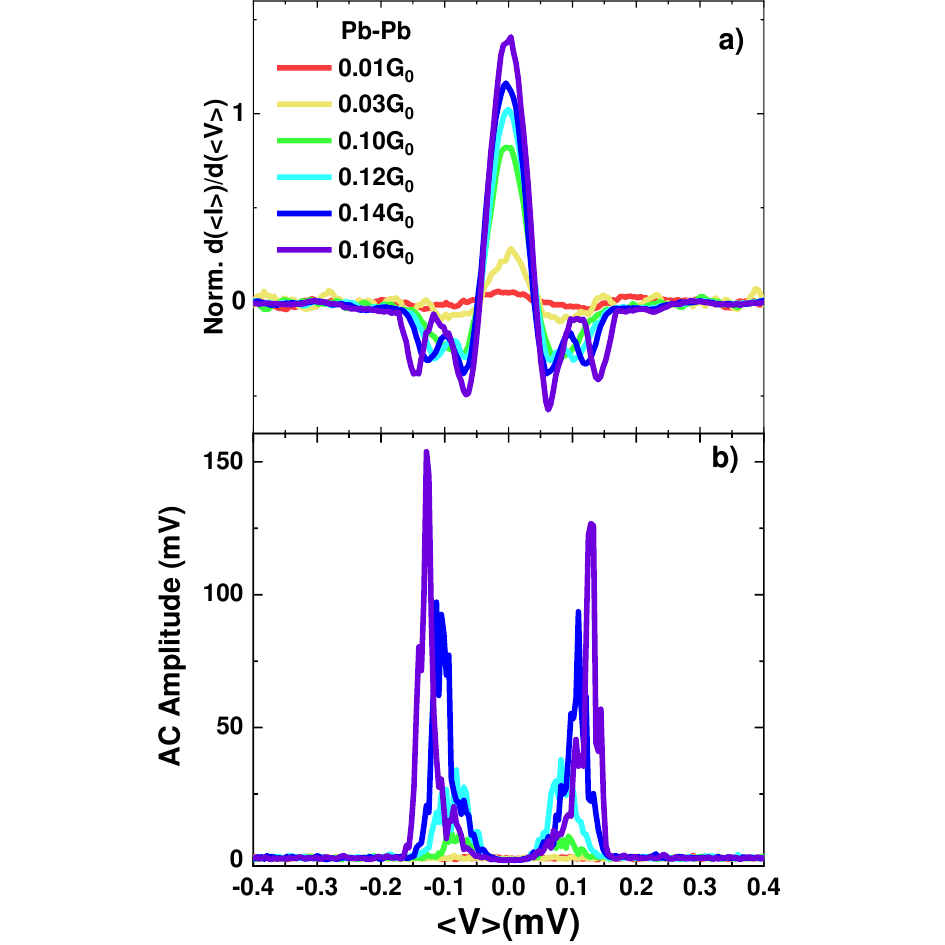}
	\end{center}
	\vskip -0.5cm
	\caption{\noindent {\bf \ | Conductance vs time-averaged voltage and magnitude of the oscillatory signal in the bistable regime.} {\bf a} Normalized conductance vs time-averaged voltage for different values of the normal state conductance (the conductance of the junction at voltages above the superconducting gap). Note the two negative resistance peaks for each sign of $\left<V\right>$. These are due to the two portions of the time averaged current vs time-averaged voltage curve (e.g. Extended Data Fig.\,8b). {\bf b} Amplitude of the spontaneous (i.e. obtained without a drive) time-dependent signal (Fig.\,3b) at the frequency of the oscillatory signal in the bistable regime (7.4 kHz here) vs the time-averaged voltage $\left<V\right>$, measured directly at the output of the current-voltage operational amplifier (Extended Data Fig.\,5), without further amplification.}
	\label{LockIn}
\end{figure}

\end{document}